\documentclass[12pt,preprint]{aastex}

\slugcomment{}

\shorttitle{}
\shortauthors{Tombesi et al.}

\begin{document}


\title{Discovery of ultra-fast outflows in a sample of Broad Line Radio Galaxies observed with Suzaku}


\author{F. Tombesi$^{1,2,3,4}$, R. M. Sambruna$^{3,4}$, J. N. Reeves$^5$, V. Braito$^6$, L. Ballo$^7$, J. Gofford$^5$, M. Cappi$^2$ and R. F. Mushotzky$^3$}
\affil{$^1$ Dipartimento di Astronomia, Universit\`a di Bologna, Via Ranzani 1, I-40127 Bologna, Italy; tombesi@iasfbo.inaf.it}
\affil{$^2$ INAF-IASF Bologna, Via Gobetti 101, I-40129 Bologna, Italy}
\affil{$^3$ Laboratory for High Energy Astrophysics, NASA/Goddard Space Flight Center, Greenbelt, MD 20771, USA}
\affil{$^4$ Department of Physics and Astronomy, George Mason University, 4400 University Drive, Fairfax, VA 22030, USA}
\affil{$^5$ Astrophysics Group, School of Physical and Geographical Sciences, Keele University, Keele, Staffordshire ST5 5BG, UK}
\affil{$^6$ Department of Physics and Astronomy, University of Leicester, University Road, Leicester LE1 7RH, UK}
\affil{$^7$ Istituto de Fisica de Cantabria (CSIC-UC), 39005 Santander, Spain}




\begin{abstract}

We present the results of a uniform and systematic search for
blue-shifted Fe K absorption lines in the X-ray spectra of five bright
Broad-Line Radio Galaxies (BLRGs) observed with Suzaku.  We detect,
for the first time at X-rays in radio-loud AGN, several absorption
lines at energies greater than 7~keV in three out of five sources,
namely 3C~111, 3C~120 and 3C~390.3.  The lines are detected with high
significance according to both the F-test and extensive Monte Carlo
simulations.  Their likely interpretation as blue-shifted Fe XXV and
Fe XXVI K-shell resonance lines implies an origin from highly ionized
gas outflowing with mildly relativistic velocities, in the range
$v$$\simeq$0.04--0.15c.  A fit with specific photo-ionization models
gives ionization parameters in the range
log$\xi$$\simeq$4--5.6~erg~s$^{-1}$~cm and column densities of
$N_H$$\simeq$$10^{22}$--$10^{23}$~cm$^{-2}$.  These characteristics
are very similar to those of the Ultra-Fast Outflows (UFOs) previously
observed in radio-quiet AGN.  Their estimated location within
$\sim$0.01--0.3pc from the central super-massive black hole suggests a
likely origin related with accretion disk winds/outflows.  Depending
on the absorber covering fraction, the mass outflow rate of these UFOs
can be comparable to the accretion rate and their kinetic power can
correspond to a significant fraction of the bolometric luminosity and
is comparable to their typical jet power.  Therefore, these UFOs can
play a significant role in the expected feedback from the AGN on the
surrounding environment and can give us further clues on the relation 
between the accretion disk and the formation of winds/jets in 
both radio-quiet and radio-loud AGN.

\end{abstract}





\section{Introduction}

Absorption from layers of photo-ionized gas in the circumnuclear
regions of AGN is commonly observed in more than half of radio-quiet
objects, the so-called warm absorbers (e.g., Blustin et al.~2005;
McKernan et al.~2007). These absorbers are usually detected in the
X-ray spectra at energies below $\sim$2--3~keV.  The typical
characteristics of this material are an ionization parameter of
log$\xi\sim$0--2~erg~s$^{-1}$~cm, a column density of
$N_H$$\sim$$10^{20}$--$10^{22}$~cm$^{-2}$ and an outflow velocity of
$\sim$100--1000~km/s. It has been suggested that the origin of this gas
might be connected with the Optical-UV Broad Line Region or with torus
winds (e.g., Blustin et al.~2005; McKernan et al.~2007).

In addition, there have been several papers in the literature recently
reporting the detection of blue-shifted Fe K absorption lines at
rest-frame energies of $\sim$7--10~keV in the X-ray spectra of
radio-quiet AGN (e.g., Chartas et al.~2002; Chartas et al.~2003; Pounds
et al.~2003; Dadina et al.~2005; Markowitz et al.~2006; Braito et
al.~2007; Cappi et al.~2009; Reeves et al.~2009a).  These lines are
commonly interpreted as due to resonant absorption from Fe XXV and/or
Fe XXVI associated with a zone of circumnuclear gas photo-ionized by the
central X-ray source, with ionization parameter
log$\xi$$\sim$3--5~erg~s$^{-1}$~cm and column density
$N_H$$\sim$$10^{22}$--$10^{24}$~cm$^{-2}$.  The energies of these
absorption lines are systematically blue-shifted and the corresponding
velocities can reach up to mildly relativistic values of
$\sim$0.2--0.4c.  In particular, a uniform and systematic search for
blue-shifted Fe K absorption lines in a large sample of radio-quiet
AGN observed with XMM-Newton has been performed by Tombesi et
al.~(2010).  This allowed the authors to assess their global detection
significance and to overcome any possible publication bias
(e.g., Vaughan \& Uttley 2008).  The lines were detected in $\sim$40\%
of the objects, and are systematically blue-shifted implying large
outflow velocities, even larger than 0.1c in $\sim$25\% of the sources.  These
findings, corroborated by the observation of short time-scale
variability ($\sim$100~ks), indicate that the absorbing material is
outflowing from the nuclear regions of AGN, at distances of the
order of $\sim$100~$r_s$ (Schwarzschild radii,
$r_s=2GM_{\mathrm{BH}}/c^2$) from the central super-massive black hole
(e.g., Cappi et al.~2009 and references therein).  Therefore, these
findings suggest the presence of previously unknown Ultra-fast
Outflows (UFOs) from the central regions of radio-quiet AGN, possibly
connected with accretion disk winds/ejecta (e.g., King \& Pounds 2003;
Proga \& Kallman 2004; Ohsuga et al.~2009; King 2010) or the base of a
possible weak jet (see the ``aborted jet'' model by Ghisellini et
al.~2004).  The mass outflow rate of these UFOs can be comparable to
the accretion rate and their kinetic energy can correspond to a
significant fraction of the bolometric luminosity (e.g., Pounds et
al.~2003; Dadina et al.~2005; Markowitz et al.~2006; Braito et
al.~2007; Cappi et al.~2009; Reeves et al.~2009a). Therefore, they
have the possibility to bring outward a significant amount of mass and
energy, which can have an important influence on the surrounding
environment (e.g., see review by Cappi 2006).  In fact, feedback from
the AGN is expected to have a significant role in the evolution of the
host galaxy, such as the enrichment of the ISM or the reduction of star
formation, and could also explain some fundamental relations (e.g., see
review by Elvis 2006 and Fabian 2009).  Moreover, the ejection of a
substantial amount of mass from the central regions of AGN can also
inhibit the growth of the super-massive black holes (SMBHs),
potentially affecting their evolution.  The study of UFOs can
also give us further clues on the relation between the accretion disk
and the formation of winds/jets.

Evidence for winds/outflows in radio-loud AGN in the X-rays has been
missing so far.  However, thanks to the superior sensitivity and
energy resolution of current X-ray detectors, we are now beginning to
find evidence for outflowing gas in radio-loud AGN as well. In fact,
the recent detection of a warm absorber in the Broad-Line Radio Galaxy
(BLRG) 3C~382 (Torresi et al.~2010; Reeves et al. 2009b) has been the
starting point for a change to the classical picture of the
radio-quiet vs. radio-loud dichotomy, at least in the X-ray domain.
This gas has an ionization parameter of
log$\xi\simeq$2--3~erg~s$^{-1}$~cm, a column density of
$N_H$$\simeq$$10^{21}$--$10^{22}$~cm$^{-2}$ and is outflowing with a
velocity of $\sim$800--1000~km/s. These parameters are somewhat
similar to those of the typical warm absorbers of Seyfert 1 galaxies
(e.g., Blustin et al.~2005; McKernan et al.~2007), which are the
radio-quiet counterparts of BLRGs.  This result indicates the presence
of ionized outflowing gas in a radio-loud AGN at a distance of
$\sim$100~pc from the central engine, suggesting its possible
association with the Optical-UV Narrow Line Region (Torresi et
al.~2010; Reeves et al.~2009b).

In this paper, we present the detection, for the first time, of
ionized ultra-fast outflows in BLRGs {\it on sub-pc scales} from
Suzaku observations. The sources in the sample -- 3C~111, 3C~390.3,
3C~120, 3C~382, and 3C~445 -- were observed with Suzaku by us as part
of our ongoing systematic study of the X-ray properties of BLRGs
(Sambruna et al. 2009), with the exception of 3C~120 which was
observed during the Guaranteed Time Observer period (Kataoka et
al. 2007).  These five BLRGs represent the ``classical'' X-ray
brightest radio-loud AGN, well studied at X-rays with previous
observatories.  Thanks to the high sensitivity of the XIS
detectors and the long net exposures of these observations of
$\sim$100~ks, we have been able to reach a high S/N in the Fe K band
that allowed, for the first time, to obtain evidence for UFOs in these
sources, in the form of blue-shifted Fe~K absorption lines at energies
greater than 7~keV.  The presence of UFOs in radio-loud AGN provides a
confirmation of models for jet-disk coupling and stresses the
importance of this class of sources for AGN feedback mechanisms. Full
accounts of the broad-band Suzaku spectra for each source will be
given in forthcoming papers.

This paper is structured as follows. In \S~2 and \S~3 we describe the
Suzaku data reduction and analysis, including statistical tests used
to assess the reality of the Fe~K absorption features (\S~3.3) and
detailed photo-ionization models used for the fits (\S~3.4). The
general results are given in \S~4, while \S~5 presents the Discussion
with the Conclusions following in \S~6. Appendix A contains the
details of the spectral fits for each BLRG and Appendix B a consistency 
check of the results. Throughout this paper, a
concordance cosmology with H$_0=71$ km s$^{-1}$ Mpc$^{-1}$,
$\Omega_{\Lambda}$=0.73, and $\Omega_m$=0.27 (Spergel et al. 2003) is
adopted. The power-law spectral index, $\alpha$, is defined such that
$F_{\nu} \propto \nu^{-\alpha}$. The photon index is $\Gamma=\alpha+1$.

\section{Suzaku observations and data reduction}

The observational details for the five BLRGs observed with Suzaku
(Mitsuda et al.~2007) are summarized in Table~1. The data were taken
from the X-ray Imaging Spectrometer (XIS, Koyama et al.~2007) and 
processed using v2
of the Suzaku pipeline. The observations were taken with the XIS
nominal (on-axis) pointing position, with the exception of the 3C~111
observation, which was taken with HXD nominal pointing.  The Suzaku
observation of 3C~120 is composed by four different exposures of
$\sim$40~ks each, taken over a period of about one month (see
Table~1).  We looked at the individual spectra and found that while
observation 2, 3 and 4 did not change significantly overall,
observation 1 instead showed a stronger X-ray emission, especially in
the soft X-ray part of the spectrum, in agreement with Kataoka et
al. (2007).  Therefore, we decided to add only observations 2, 3 and 4
(we will call this observation 3C~120b) and to analyze the spectrum of
observation 1 separately (we will call this observation 3C~120a).

Data were excluded within 436 seconds of passage through the South
Atlantic Anomaly (SAA) and within Earth elevation angles or Bright
Earth angles of $<5^\circ$ and $<20^\circ$, respectively. XIS data were
selected in $3 \times 3$ and $5 \times 5$ edit-modes using grades
0, 2, 3, 4, 6, while hot and flickering pixels were removed using the {\sc
  sisclean} script. Spectra were extracted from within circular
regions of between 2.5\arcmin--3.0\arcmin\, radius, while background
spectra were extracted from circles offset from the source and
avoiding the chip corners containing the calibration sources. The
response matrix ({\sc rmf}) and ancillary response ({\sc arf}) files
were created using the tasks {\sc xisrmfgen} and {\sc xissimarfgen},
respectively, the former accounting for the CCD charge injection and
the latter for the hydrocarbon contamination on the optical blocking
filter. 

Spectra from the front illuminated XIS~0, XIS~2 (where available) and
XIS~3 chips were combined to create a single source spectrum
(hereafter XIS-FI). Given its superior sensitivity in the region of
interest, 3.5--10.5~keV, we restricted our analysis to the XIS-FI
data.  The data from the back illuminated XIS~1 (hereafter XIS-BI)
chip were analysed separately and checked for consistency with the
XIS-FI results.  In all cases, the power-law continuum and Fe
K$\alpha$ emission line parameters are completely consistent. Instead,
the lower S/N of the XIS-BI in the 4--10~keV band ($\sim$40\% of the
XIS-FI) allowed us to place only lower limits to the equivalent width
of the blue-shifted absorption lines (see Appendix B and Table~5).

Furthermore, Appendix B gives more details on the various consistency
checks we have performed in order to verify the reality of the
absorption lines detected in the 7--10~keV band. In particular, we 
determined that the XIS background has a negligible effect on the
detection of each of the individual absorption lines and we checked the 
consistency of the results among the individual XIS cameras (see Table~5). 
We also tested that the alternative modeling of the lines with ionized Ni 
K-shell transitions and ionized Fe K edges is not feasible. Finally, in 
\S4.2 we verified the fit results from the broad-band (E$=$0.5--50~keV) 
XIS$+$PIN spectra.

\section{Spectral fits}

We performed a uniform spectral analysis of the small sample of five BLRGs
in the Fe K band (E$=$3.5--10.5~keV). We used the \emph{heasoft}
v. 6.5.1 package and XSPEC v. 11.3.2.  We extracted the source spectra
for all the observations, subtracted the corresponding background and
grouped the data to a minimum of 25 counts per energy bin to enable
the use of the $\chi^2$ when performing spectral fitting. Fits were limited
to the 3.5--10.5~keV energy band.

\subsection{The baseline model}

As plausible phenomenological representation of the continuum in
3.5--10.5~keV, we adopt a single power-law model.  
We did not find it necessary
to include neutral absorption from our own Galaxy as the
relatively low column densities involved
(Dickey \& Lockman 1990; Kalberla et al.~2005) have negligible effects in
the considered energy band, see Table~1.
The only exception is 3C~445, where the
continuum is intrinsically absorbed by a column density of
neutral/mildly-ionized gas as high as $N_H$$\sim$$10^{23}$~cm$^{-2}$
(Sambruna, Reeves \& Braito 2007); for this source we included also a
neutral intrinsic absorption component with a column density of
$N_H$$\simeq$$2\times 10^{23}$~cm$^{-2}$ (see Table~2). A more
detailed discussion of absorption in this source using Chandra and
Suzaku data is presented in Reeves et al. (2010, in prep.) and
Braito et al. (2010, in prep.).

The ratios of the spectral data against the simple (absorbed for 3C~445) 
power-law continuum
for the five BLRGs are shown in the upper panels of Fig.~1, Fig.~3 and
Fig.~5. Some additional spectral complexity can be clearly seen, such
as an ubiquitous, prominent neutral Fe K$\alpha$ emission line at the
rest frame energy of 6.4~keV, absorption structures at energies
greater than 7~keV (3C~111, 3C~120, and 3C~390.3) and narrow emission
features red-ward (3C~445) and blue-ward (3C~120 and 3C~382) to the
neutral Fe K$\alpha$ line. To model the emission lines we added
Gaussian components to the power law model, including the Fe K$\alpha$
emission line at E$\simeq$6.4~keV and ionized Fe~K emission lines in
the energy range E$\sim$6.4--7~keV, depending on the ionization state
of iron, which in this energy interval is expected to range from Fe~II
up to Fe~XXVI.

We find that the baseline model composed by a power-law plus Gaussian Fe K 
emission lines provides an excellent phenomenological
characterization of the 3.5--10.5~keV XIS data with the lowest number
of free parameters.  The results of the fits for the five BLRGs are
reported in Table~2.  Note that only those emission lines with detection
confidence levels greater than 99\% were retained in following
fits. The weak red-shifted emission line present in 3C~445 was not
included because this has negligible effect on the fit results; this 
line will be discussed by Braito et al. (2010 in prep.).

\subsection{Fe K absorption lines search}

As apparent from Figure~1, 3, and 5, several absorption dips are
present in the residuals of the baseline model in various cases. To
quantify their significance, we computed the $\Delta\chi^2$ deviations
with respect to the baseline model (\S~3.1) over the whole 3.5--10.5~keV
interval. The method is similar to the one used by the \emph{steppar}
command in XSPEC to visualize the error contours, but in this case the
inner contours indicate higher significance than the outer ones
(e.g., Miniutti \& Fabian 2006; Miniutti et al.~2007; Cappi et
al.~2009; Tombesi et al.~2010).

The analysis has been carried out for each source spectrum as follows:
1) we first fitted the 3.5--10.5~keV data with the baseline model and
stored the resulting $\chi^2$; 2) a further narrow, unresolved
($\sigma=$10~eV) Gaussian test line was then added to the model, with
its normalization free to have positive or negative values. Its
energy was stepped in the 4--10~keV band at intervals of
100~eV in order to properly sample the XIS energy resolution, each
time making a fit and storing the resulting $\chi^2$ value. In this
way we derived a grid of $\chi^2$ values and then plot the contours
with the same $\Delta\chi^2$ with respect to the baseline model.

The contour plots for the different sources are reported in the lower
panel of Figures~1, 3 and 5.  The contours refer to $\Delta\chi^2$
levels of $-2.3$, $-4.61$ and $-9.21$, which correspond to F-test
confidence levels of 68\% (red), 90\% (green) and 99\% (blue),
respectively. The position of the neutral Fe K$\alpha$ emission line
at rest-frame energy E$=$6.4~keV is marked by the dotted vertical
line. The arrows indicate the position of the blue-shifted absorption
lines detected at $\ge$99\%. The black contours indicate the baseline
model reference level ($\Delta\chi^2=+0.5$).

We then proceeded to directly fit the spectra, adding Gaussian
absorption lines where indications for line-like absorption features
with confidence levels greater than 99\% were present. As already
noted in \S3.1, we checked that neglecting to include the weak
red-shifted emission line apparent only in the spectrum of 3C~445 has
no effect in the fit results. The detailed fitting and modeling of the
Fe K absorption lines is reported in Table~3 and is discussed in the
Appendix A for each source.

\subsection{Line significance from Monte Carlo simulations}

The contour plots in the lower panels of Fig.~1, Fig.~3 and Fig.~5
visualize the presence of spectral structures in the data and
simultaneously give an idea of their energy, intensity and confidence
levels using the standard F-test. However, they give only a
semi-quantitative indication and the detection of each line must be
confirmed by directly fitting the spectra.  Moreover, it has been
demonstrated that the F-test method can slightly overestimate the
actual detection significance for a blind search of
emission/absorption lines as it does not take into account the
possible range of energies where a line might be expected to occur,
nor does it take into account the number of bins (resolution elements)
present over that energy range (e.g., Protassov et al.~2002).  This
problem requires an additional test on the red/blue-shifted lines
significance and can be solved by determining the unknown underlying
statistical distribution by performing extensive Monte Carlo (MC)
simulations (e.g., Porquet et al.~2004; Yaqoob \& Serlemitsos 2005;
Miniutti \& Fabian 2006; Markowitz et al.~2006; Cappi et al.~2009;
Tombesi et al.~2010).

Therefore, we performed detailed MC simulations to estimate
the actual significance of the absorption lines detected at energies
greater than 7~keV.  We essentially tested the null hypothesis that
the spectra were adequately fitted by a model that did not include the
absorption lines.  The simulations have been carried out as follows:
1) we simulated a source spectrum using the \emph{fakeit} command in
XSPEC and assuming the baseline model listed in Table~2 without any
absorption lines and with the same exposure as the real data. We
subtracted the appropriate background and grouped the data to a
minimum of 25 counts per energy bin; 2) we fitted the faked spectrum
with the baseline model in the 3.5--10.5~keV band, stored the new
parameters values and generated another simulated spectrum as in step
2 but using the refined model. This procedure accounts for the
uncertainty in the null hypothesis model itself and is particularly
relevant when the original data set is noisy; 3) the newly simulated
spectrum was fitted again with the baseline model in the 3.5--10.5~keV
and the resultant $\chi^2$ was stored; 4) then, a further Gaussian
line (unresolved, $\sigma=$10~eV) was added to the model, with its
normalization initially set to zero and let free to vary between
positive and negative values. To account for the range of energies in
which the line could be detected in a blind search, we stepped its
centroid energy between 7~keV and 10~keV at intervals of 100~eV to
sample the XIS energy resolution, fitting each time and storing
only the maximum of the resultant $\Delta\chi^2$ values. The procedure
was repeated $S=1000$ times and consequently a distribution of
simulated $\Delta\chi^2$ values was generated. The latter indicates
the fraction of random generated emission/absorption features in the
7--10~keV band that are expected to have a $\Delta\chi^2$ greater than
a threshold value. In particular, if $N$ of the simulated
$\Delta\chi^2$ values are greater or equal to the real value, then the
estimated detection confidence level from MC simulations is
simply $1-N/S$. 

The MC detection probabilities for the absorption lines are given in 
Table~3. The values are in the range between 91\% and 99.9\%.
As expected, these estimates are slightly lower than those derived from the 
F-test ($\ge$99\%) because they effectively take into account the random 
generated lines in the whole 7--10~keV energy interval.

\subsection{Photo-ionization modeling}

To model the absorbing material that is photo-ionized by the nuclear
radiation, a grid with the Xstar code (Kallman \& Bautista 2001) was
generated. We modeled the nuclear X-ray ionizing continuum with a
power-law with photon index $\Gamma=2$, as usually assumed for Seyfert
galaxies, which takes into account the possible steeper soft
excess component (e.g., Bianchi et al.~2005).  A different choice of
the power-law slope in the range $\Gamma$$=$1.5--2.5 has negligible
effects ($<$5\%) on the parameter estimates in the considered Fe K
band, E$=$3.5--10.5~keV.  Moreover, as already noted by McKernan et
al.~(2003a), the presence or absence of the possible UV-bump in the
SED has a negligible effect on the parameters of the photo-ionized gas
in the Fe K band because in this case the main driver is the ionizing
continuum in the hard X-rays (E$>$6~keV).  Standard solar abundances
are assumed throughout (Grevesse et al.~1996).

The velocity broadening of absorption lines from the photo-ionized
absorbers in the central regions of Seyfert galaxies is dominated by
the turbulence velocity component, commonly assumed to be in the range
$\sim$100--1000~km/s (e.g., Bianchi et al.~2005; Risaliti et al.~2005;
Cappi et al.~2009 and references therein).  The energy resolution of
the XIS instruments in the Fe K band is FWHM$\sim$100--200~eV,
implying that lines with velocity broadening lower than
$\sim$2000--4000~km/s are unresolved.  Therefore, given that we cannot
estimate the velocity broadening of the lines directly from the
spectral data, we generated an Xstar grid assuming the most likely
value for the turbulent velocity of the gas of 500~km/s.  We checked
that for higher choices of this parameter, the resultant
estimate of the ionization parameter was not affected, although the
derived absorber column density was found to be slightly lower.  
This is due to the fact that the core of the line tends to
saturate at higher $N_H$, with increasing the velocity broadening 
(e.g., Bianchi et al.~2005). The opposite happens for 
lower choices of the turbulent velocity. However, the
resulting difference of $\sim$5--10\% in the derived values is
completely negligible and well within the measurement errors.

Therefore, we apply this photo-ionization grid to model directly the
different absorption lines detected in the Fe K band. 
The free parameters of the model are the absorber column density 
$N_H$, the ionization parameter $\xi$ and the velocity shift $v$. We let the 
code find the best-fit values and it turned out that the gas is 
systematically outflowing, with velocities consistent with those derived from 
the Gaussian absorption lines fits (see \S3.2 and Appendix A for a detailed discussion of each source). 
The Xstar parameters are reported in Table~4 and the best-fit models are shown 
in Fig.~2 and Fig.~4. A consistency check of the results from a broad-band spectral analysis is reported in \S4.2.

\section{General results}

\subsection{Fe K band spectral analysis}

In this Section we summarize the results of the spectral fits to the
3.5--10.5~keV XIS-FI spectra of the BLRGs of our sample with a model
consisting of the baseline model plus absorption lines and a detailed
photoionization grid (see above). Results for individual sources are
discussed in Appendix A. 

The results of the fits with the baseline model are listed in Table~2,
while the residuals of this model are shown in Figures~1, 3, and 5 for
the five BLRGs, together with the $\Delta\chi^2$ contours. As
mentioned above, absorption dips are visible and to assess their
statistical significance we used both the F-test and extensive Monte
Carlo simulations. The results of these tests, reported in Table~3,
establish that only in 3/5 sources we do detect reliably absorption
features at energies $\sim$ 7.3--7.5~keV and 8.1--8.7~keV, namely in
3C~111, 3C~120b, and 3C~390.3. In these three sources, the absorption
lines are detected with confidence levels higher than 99\% with the F-test
and higher than 91\% with the Monte Carlo method (Table~3). We fitted the
absorption features by adding narrow Gaussian components, or a blend
of narrow components, to the baseline model. The Gaussian parameters
are reported in Table~3.

Given the high cosmic abundance of Fe, the most intense spectral
features expected from a highly ionized absorber in the 3.5--10.5~keV
band are the K-shell resonances of Fe XXV and Fe XXVI (e.g., Kallman
et al.~2004).  However, the rest-frame energies of the detected
absorption lines are in the range $\simeq$7.3--7.5~keV and
$\simeq$8.1--8.8~keV, larger than the expected energies of the atomic
transitions for Fe XXV and Fe XXVI. An interesting possibility is that
the absorption lines detected in the BLRGs are due, similarly to those
recently observed in Seyferts, to blueshifted resonant lines of highly
ionized Fe, thus implying the presence of fast outflows in radio-loud AGN
as well. If we hold this interpretation true, the derived outflow
velocities are in the range $\simeq$0.04--0.15c. 

We also performed more physically consistent spectral fits using the
Xstar photoionization grid described in \S~3.4 (see Fig.~2 and
Fig.~4). Good fits are obtained with this model, yielding ionization
parameters log$\xi$$\simeq$4--5.6~erg~s$^{-1}$~cm and column densities
 $N_H$$\simeq$$10^{22}$--$10^{23}$~cm$^{-2}$.  The
derived blue-shifted velocities are consistent with those from the
simple phenomenological fits, $v$$\simeq$0.04--0.15c (see Table~4). We
note that, given the very high ionization level of this absorbing
material, no other significant signatures are expected at lower
energies as all the elements lighter than iron are almost completely
ionized. 

An important caveat is that the velocities and column densities
derived by fitting the spectral data with the Xstar grid depend on the
unknown inclination angle of the outflow with respect to the line of
sight.  In other words, they depend on whether we are actually looking
directly down to the outflowing stream or intercepting only part of
it (e.g., Elvis 2000).  Therefore, the obtained values (see Table~4)
are only conservative estimates and represent lower limits. 

In conclusion, we detected for the first time in radio-loud AGN at
X-rays, absorption lines in the energy range 7--10~keV in the Suzaku
XIS spectra of 3/5 BLRGs -- 3C~111, 3C~390.3, and 3C~120. If
interpreted as blueshifted resonant absorption lines of highly ionized
Fe, the features imply the presence of ultra-fast (v $\sim$
0.04-0.15c) outflows in the central regions of BLRGs. In \S5 we discuss
more in depth this association and the inferred outflow physical
properties.

\subsection{Broad-band spectral analysis}

As a consistency check of the Fe K band (E$=$3.5--10.5~keV) based results, we exploited the broad-band capabilities of Suzaku combining the XIS and PIN spectra.
The energy band covered in this way is very broad, from 0.5~keV up to 50~keV.
We downloaded and reduced the PIN data of 3C~111, 3C~390.3 and 3C~120 and analyzed the combined XIS-FI and PIN spectra.
For 3C~390.3 and 3C~120  we applied the broad-band models already published in the literature by Sambruna et al.~(2009) and Kataoka et al.~(2007). 
Instead, for 3C~111, we used the broad-band model that will be reported by us in Ballo et al.~(2010, in prep.). This is essentially composed by a power-law continuum with Galactic absorption, plus cold reflection ($R\la$1) and the Fe K$\alpha$ emission line at E$\simeq$6.4~keV. The resultant power-law photon index of this fit is $\Gamma$$\simeq$1.6, which is slightly steeper than the estimate of $\Gamma$$\simeq$1.5 from the local continuum in the 3.5--10.5~keV band (see Table~2). We included the neutral Galactic absorption component in all broad-band fits (see Table~1). 
Then, we modeled the blue-shifted absorption lines with the Xstar photo-ionization grid already discussed in \S3.4., letting the column density, the ionization parameter and the velocity shift vary as free parameters.

The best-fit estimates of the Fe K absorbers derived from these broad-band fits are completely consistent with those reported in Table~4.
In particular, for 3C~111 we obtain an ionization parameter of log$\xi$$=$$4.9^{+0.2}_{-0.4}$~erg~s$^{-1}$~cm, a column density of $N_H$$>$$1.5\times 10^{23}$~cm$^{-2}$ and an outflow velocity of $v_{out}$$=$$+0.039\pm0.003$c.
For 3C~390.3, we estimate log$\xi$$=$$5.6\pm0.5$~erg~s$^{-1}$~cm, $N_H$$>$$2\times 10^{22}$~cm$^{-2}$ and $v_{out}$$=$$+0.146\pm0.007$c. 
Finally, for 3C~120b, we derive log$\xi$$=$$3.7\pm0.2$~erg~s$^{-1}$~cm, $N_H$$=$$(1.5\pm0.4)\times 10^{22}$~cm$^{-2}$ and $v_{out}$$=$$+0.075\pm0.003$c.
This also assures that the addition of a weak reflection component with $R$$<$1 (e.g., Sambruna et al.~2009; Kataoka et al.~2007; Ballo et al.~2010 in prep.) does not change at all the fit results.

Moreover, it is important to note here that we do not find any evidence for a lower ionization (log$\xi$$\la$3~erg~s$^{-1}$~cm) warm absorber at E$\la$3~keV in these three sources. This rules out any possible systematic contamination from moderately ionized iron and strengthens the interpretation of the absorption lines at E$>$7~keV as genuine blue-shifted Fe XXV and Fe XXVI K-shell transitions.
As already introduced in \S1, the only object with the detection of a soft X-ray warm absorber in its high energy resolution Chandra HETG (Reeves et al.~2009b) and XMM-Newton RGS (Torresi et al.~2010) spectra is 3C~382.  On the other hand, a heavy soft X-ray absorption from neutral/mildly-ionized gas with $N_H$$\sim$$10^{23}$~cm$^{-2}$ has been reported in the XMM-Newton spectrum of 3C~445 (Sambruna, Reeves \& Braito 2007). This result is also confirmed by a Chandra LETG and a Suzaku broad-band spectral analysis presented in Reeves et al.~(2010, in prep.) and Braito et al.~(2010, in prep.), respectively. 
However, we did not find any significant narrow Fe K absorption 
line features in the 7--10~keV Suzaku XIS spectra of these two sources 
from this analysis.

\section{Discussion}

\subsection{Evidence for Ultra-fast Outflows in BLRGs} 

The discovery of Ultra-Fast Outflows (UFOs) in radio-loud BLRGs
parallels the detection of UFOs in the X-ray spectra of several
Seyfert galaxies and radio-quiet quasars (e.g., Chartas et al.~2002;
Chartas et al.~2003; Pounds et al.~2003; Dadina et al.~2005; Markowitz
et al.~2006; Braito et al.~2007; Cappi et al.~2009; Reeves et
al.~2009a).  The presence of UFOs in radio-quiet sources was recently
established through a systematic, uniform analysis of the XMM-Newton archive
on a large number of sources (Tombesi et al. 2010), overcoming
possible publication biases (e.g., Vaughan \& Uttley 2008).

While a uniform analysis was also performed in this work, it should be
noted that our small sample is not complete and the results might not
be representative of the global population of BLRGs.  Therefore, to
obtain better constraints on the statistical incidence and parameters
of UFOs in BLRGs, it is imperative to expand the sample of sources
with high-quality X-ray observations in the next few years through
Suzaku and XMM-Newton observations of additional sources. 

However, it has been claimed that part (or even all) of the blue-shifted
ionized absorption features detected in the X-ray spectra of bright
AGN could be affected by contamination from local ($z\simeq0$)
absorption in our Galaxy or by the Warm/Hot Intergalactic Medium
(WHIM) at intermediate red-shifts, due to the fact that some of them
have blue-shifted velocities comparable to the sources cosmological
red-shifts (e.g., McKernan et al.~2003b; McKernan et al.~2004; McKernan
et al.~2005).  We performed some tests to look into this scenario.  We
can use the velocity information and compare the absorber blue-shifted
velocities with the cosmological red-shifts of the sources. The
blue-shifted velocities of the absorbers detected in 3C~120 and
3C~390.3 (see Table~4) are much larger than the sources cosmological
red-shifts (see Table~1). This conclusion is strong enough to rule out
any contamination due to absorption from local or intermediate
red-shift material in these two sources.  However, the derived
blue-shifted velocity of $v$$=$$+0.041\pm0.003$c for the highly
ionized absorber in 3C~111 is instead somewhat similar to the source
cosmological red-shift of $z=0.0485$ and needs to be investigated in
more detail.  The difference between the two values is
$zc-v$~$\simeq$~$0.007$c, which could indicate absorption from highly
ionized material either in our Galaxy and outflowing with that
velocity ($v$$\sim$2000~km/s) along the line of sight or at rest and
located at that intermediate red-shift ($z$$\simeq$$0.007$).

The galaxy 3C~111 is located at a relatively low latitude ($b=
-8.8\degr$) with respect to the Galactic plane and therefore its X-ray
spectrum could be, at some level, affected by local
obscuration. However, the estimated column density of Galactic
material along the line of sight of the source is $N_H\sim 3\times
10^{21}$~cm$^{-2}$ (Dickey \& Lockman 1990; Kalberla et al.~2005),
which is far too low to explain the value of $N_H\sim
10^{23}$~cm$^{-2}$ of the absorber from fits of the Suzaku spectrum
(see Table~4). Nevertheless, the source is located near the direction of the 
Taurus molecular cloud, which is the nearest large star-forming region in our
Galaxy.

A detailed optical and radio study of the cloud has been reported by
Ungerer et al.~(1985).  From the analysis of the emission from the
stars in that region and the molecular emission lines, the authors 
estimated several parameters of the cloud, such
as the location at a distance of $\sim$400~pc (with a linear extent of
$\sim$5~pc), a kinetic temperature of $T\simeq$10~K, a typical
velocity dispersion of $\sim$1--3~km/s and a low number density of
$n(\mathrm{H_2})$$\sim$300~cm$^{-3}$.  These parameters are completely
inconsistent with the properties of the X-ray absorber.  In fact, the
extreme ionization level ($\log\xi\sim5$~erg~s$^{-1}$~cm) needed to
have sufficient Fe XXVI ions would completely destroy all the
molecules and ionize all the lighter atoms. The temperature associated
with such photo-ionized absorber ($T$$\sim$$10^6$~K) is larger than
estimated for the Taurus molecular cloud. Also the outflow
velocity of $\sim$2000~km/s, expected if associated with such Galactic
clouds, would be substantially higher than the velocity dispersion
estimated by Ungerer et al.~(1985).  The authors also stated that the
mapping of the visual extinction due to the molecular cloud clearly
shows that the region of the cloud in front of 3C~111 is not the
densest part (see Fig.~3 of Ungerer et al.~1985).

This result is also supported by a recent detailed X-ray study of this
region that has been performed by the XMM-Newton Extended Survey of
the Taurus Molecular Cloud project (G\"udel et al.~2007).  This work
has been focused on the study of the stars and gas located in the most
populated $\simeq$5 square degrees region of the Taurus cloud.  With a
declination of $\sim$$38\degr$, 3C~111 is located outside the edge of
this complex region, where mainly only extended cold and low density
molecular clouds are distributed (see Fig.~1 of G\"udel et al.~2007).
Therefore, the identification of the highly ionized absorber of 3C~111
with local Galactic absorption is not feasible.

We also find that association with absorption from the WHIM at intermediate 
red-shift ($z$$\sim$$0.007$) is very unlikely. In fact, this diffuse gas is 
expected to be
collisionally ionized, instead of being photo-ionized by the AGN
continuum. Therefore, the temperature required to have a substantial
He/H-like iron population would be much higher
($T$$\sim$$10^{7}$--$10^{8}$~K) than the expected
$T$$\sim$$10^5$--$10^6$~K. The huge column density of gas
($N_H$$\ga$$10^{23}$~cm$^{-2}$) required to reproduce the observed
features is also too high compared to those expected for the WHIM
($N_H$$\la$$10^{20}$~cm$^{-2}$).  Moreover, the detection of highly
ionized absorbers in 3C~120 and 3C~390.3 with blue-shifted velocities
substantially larger than the relative cosmological red-shifts
strongly supports the association of the absorber in 3C~111 with a UFO
intrinsic to the source.

Similar conclusions were reached by Reeves et al.~(2008) concerning
the bright quasar PG~1211+143. The X-ray spectrum of this source
showed a blue-shifted absorption line from highly ionized iron (Pounds
et al.~2003; Pounds \& Page 2006) with a blue-shifted velocity
comparable to the cosmological red-shift of the source.  This led some
authors to suggest its possible association with absorption from
intervening diffuse material at $z$$\sim$0 (e.g., McKernan et
al.~2004).  However, the detection of line variability on a time-scale
less than 4~yrs, suggesting a compact $\sim$pc scale absorber, and the
extreme parameters of the absorber, e.g.,
log$\xi$$\sim$3--4~erg~s$^{-1}$~cm and
$N_H$$\sim$$10^{22}$--$10^{23}$~cm$^{-2}$, led Reeves et al.~(2008) to
exclude such interpretation.  As pointed out by the authors, the
evidence of several other radio-quiet AGN with Fe K absorption with
associated blue-shifted velocities higher than the relative
cosmological red-shift suggested that the case of PG~1211+143 was a
mere coincidence (see also Tombesi et al.~2010).  

We conclude that the evidence for UFOs in BLRGs from Suzaku data is
indeed robust. In the next section we examine their physical properties
in detail.

\subsection{Physical properties of Ultra-fast Outflows} 

From the definition of the ionization parameter $\xi=L_{ion}/nr^2$
(Tarter, Tucker \& Salpeter 1969), where $n$ is the average absorber
number density and $L_{ion}$ is the source X-ray ionizing luminosity
integrated between 1~Ryd and 1000~Ryd (1~Ryd=13.6~eV), we can estimate
the maximum distance $r$ of the absorber from the central source.  The
column density of the gas $N_H$ is a function of the density of the
material $n$ and the shell thickness $\Delta r$: $N_H= n \Delta
r$. Making the reasonable assumption that the thickness is less than
the distance from the source $r$ and combining with the expression for
the ionization parameter, we obtain the upper limit $r<L_{ion}/\xi
N_H$.  Using the absorption corrected luminosities
$L_{ion}\simeq2.2\times 10^{44}$~erg~s$^{-1}$, $L_{ion}\simeq2.3\times
10^{44}$~erg~s$^{-1}$ and $L_{ion}\simeq 5.1\times
10^{44}$~erg~s$^{-1}$ directly estimated from the Suzaku data and the
ionization parameters and column densities listed in Table~4, we
obtain the limits of $r<2\times 10^{16}$~cm ($<$0.007~pc), $r<
10^{18}$~cm ($<$0.3~pc) and $r<4\times 10^{16}$~cm ($<$0.01~pc) for
3C~111, 3C~120 and 3C~390.3, respectively.  Using the black hole mass
estimates of $M_{\mathrm{BH}}\sim3\times 10^9$$M_{\sun}$ for 3C~111
(Marchesini et al.~2004), $M_{\mathrm{BH}}\sim5\times 10^7$$M_{\sun}$
for 3C~120 (Peterson et al.~2004) and $M_{\mathrm{BH}}\sim3\times
10^{8} M_{\sun}$ for 3C~390.3 (Marchesini et al.~2004; Peterson et
al.~2004), the previous limits on $r$ correspond to a location for the
absorber within a distance of $\sim$20~$r_s$, $\sim$7$\times
10^4$~$r_s$ and $\sim$500~$r_s$ from the super-massive black hole,
respectively.  The expected variability time-scale of the absorbers
from the light crossing time, $t$$\sim$$r/c$, is $t$$\sim$600--700~ks
($\sim$7 days) for 3C~111, $t$$\sim$1~yr for 3C~120 and
$t$$\sim$15--20~days for 3C~390.3, respectively.

A rough estimate of the escape velocity along the radial distance for
a Keplerian disk can be derived from the equation
$v_{esc}^2=2GM_{\mathrm{BH}}/r$, which can be re-written as
$v_{esc}=(r_s/r)^{1/2}c$.  Therefore, for 3C~111 the escape velocity
at the location of $\sim 20 r_s$ is $v_{esc}\sim0.2$c, which is larger
than the measured outflow velocity of $v\sim$0.041c (see Table
4). This implies that most likely the absorber is actually in the form
of a blob of material which would eventually fall back down, possibly
onto the accretion disk.  For 3C~120, the measured outflow velocity
$v\sim$0.076c (see Table~4) is equal to the escape velocity at a
distance of $\sim$200$r_s$ from the black hole. Therefore, if the
launching region is further away than this distance, the ejected blob
is likely to escape the system.  Concerning 3C~390.3, the measured
velocity of $v\sim$0.146c (see Table~4) is larger than the escape
velocity at $\sim500 r_s$ and equals that at a distance of
$\sim$50--60$r_s$. Therefore, if the blob of material has been ejected
from a location between, say, $\sim100 r_s$ and $\sim500r_s$, it has
likely enough energy to eventually leave the system.

We can get an idea of the effectiveness of the AGN in producing
outflows by comparing their luminosity with the Eddington luminosity,
$L_{Edd}$$\simeq$$1.3\times10^{38}
(M_{\mathrm{BH}}/M_{\sun})$~erg~s$^{-1}$.  Substituting the estimated
black hole mass for each source, we have
$L_{Edd}$$\simeq$$3.9\times10^{47}$~erg~s$^{-1}$ for 3C~111,
$L_{Edd}$$\simeq$$6.5\times10^{45}$~erg~s$^{-1}$ for 3C~120 and
$L_{Edd}$$\simeq$$3.9\times10^{46}$~erg~s$^{-1}$ for 3C~390.3,
respectively.  From the relation $L_{bol}$$\simeq$$10L_{ion}$
(e.g., McKernan et al.~2007), the bolometric luminosities of the
different sources are:
$L_{bol}$$\simeq$$2.2\times10^{45}$~erg~s$^{-1}$ for 3C~111,
$L_{bol}$$\simeq$$2.3\times10^{45}$~erg~s$^{-1}$ for 3C~120 and
$L_{bol}$$\simeq$$5.1\times10^{45}$~erg~s$^{-1}$ for 3C~390.3,
respectively.  The ratio $L_{bol}/L_{Edd}$ is
almost negligible for 3C~111 but it is of the order of $\sim$0.1--0.4
for 3C~120 and 3C~390.3.  These two sources are emitting closer to
their Eddington limits and therefore are more capable of producing
powerful outflows/ejecta that would eventually leave the system
(e.g., King \& Pounds 2003; King 2010).  This supports the conclusions 
from the 
estimates on the location of the ejection regions and the comparison
of the outflow velocities with respect to the escape velocities. 

Moreover, assuming a constant velocity for the outflow and the
conservation of the total mass, we can roughly estimate the mass loss
rate $\dot{M}_{out}$ associated to the fast outflows,
$\dot{M}_{out}=4\pi C r^2 n m_p v$ (e.g., Blustin et al.~2005; McKernan
et al.~2007), where $v$ is the outflow velocity, $n$ is the absorber
number density, $r$ is the radial distance, $m_p$ is the proton mass
and $C\equiv(\Omega/4\pi)$ is the covering fraction, which in turn
depends on the solid angle $\Omega$ subtended by the absorber.  From
the definition of the ionization parameter $\xi$, we obtain
$\dot{M}_{out}=4\pi C \frac{L_{ion}}{\xi} m_p v$. Substituting the
relative values, we derive estimates of $\dot{M}_{out} \sim
2\,C$~M$_{\sun}~$yr$^{-1}$, $\dot{M}_{out} \sim
17\,C$~M$_{\sun}$~yr$^{-1}$ and $\dot{M}_{out} \sim
2\,C$~M$_{\sun}$~yr$^{-1}$ for 3C~111, 3C~120 and 3C~390.3,
respectively.

The kinetic power carried by the outflows can be estimated as
$\dot{E}_K\equiv \frac{1}{2} \dot{M}_{out} v^2$, which roughly
corresponds to $\dot{E}_K \sim 4.5\times 10^{43}\,C$~erg~s$^{-1}$,
$\dot{E}_K \sim 3\times 10^{45}\,C$~erg~s$^{-1}$ and $\dot{E}_K \sim
1.2\times 10^{45}\,C$~erg~s$^{-1}$ for 3C~111, 3C~120 and 3C~390.3,
respectively. Note that, depending on the estimated covering fraction,
the kinetic power injected in these outflows can be substantial,
possibly reaching significant fractions ($\sim$0.01--0.5) of the
bolometric luminosity and can be comparable to the typical jet power
of these sources of $\sim$$10^{44}$--$10^{45}$~erg~s$^{-1}$, the
latter being the power deposited in the radio lobes (Rawlings \&
Saunders 1991).

Therefore, it is important to compare the fraction of mass that goes
into accretion of the system with respect to that which is lost
through these outflows.  Following McKernan et al.~(2007), we can
derive a simple relation for the ratio between the mass outflow rate
and the mass accretion rate, i.e. $\dot{M}_{out}/\dot{M}_{acc}\simeq
6000\, C(v_{0.1}/\xi_{100})\eta_{0.1}$, where $v_{0.1}$ is the outflow
velocity in units of 0.1c, $\xi_{100}$ is the ionization parameter in
units of 100~erg~s$^{-1}$~cm and $\eta=\eta_{0.1}\times 0.1$ is the
accretion efficiency.  Substituting the parameters with their relative
values listed in Table~4, we obtain
$\dot{M}_{out}/\dot{M}_{acc}$$\sim$$2\,C$ for 3C~111,
$\dot{M}_{out}/\dot{M}_{acc}$$\sim$$40\,C$ for 3C~120 and
$\dot{M}_{out}/\dot{M}_{acc}$$\sim$$2\,C$ for 3C~390.3, respectively.

These estimates depend on the unknown value of the covering fraction
$C$. A very rough estimate of the global covering fraction of these
absorbers can be derived from the fraction of sources of our small
sample: $C\simeq f= 3/5\sim 0.6$ (e.g., Crenshaw et al.~1999).  This
suggests that the geometrical distribution of the absorbing material
is not very collimated but large opening angles are favored.  The
rough estimate $C\sim0.6$ implies the possibility of reaching ratios
of about unity or higher between the mass outflow and accretion
rates. This means that these outflows can potentially generate
significant mass and energy losses from the system.  However, the
covering fraction crude estimate of $C\sim0.6$ has been derived from a
very small sample which is far from being complete and therefore could not be
fully representative of the whole population of BLRGs.

The physical characteristics of UFOs here derived for the three
BLRGs strongly point towards an association with winds/outflows from
the inner regions of the putative accretion disk.  In fact,
simulations of accretion disks in AGN ubiquitously predict the
generation of mass outflows. For instance, the location, geometry,
column densities, ionization and velocities of our detected UFOs are
in good agreement with the AGN accretion disk wind model of Proga \&
Kallman (2004). In this particular model the wind is driven by
radiation pressure from the accretion disk and the opacity is
essentially provided by UV lines.  Depending on the angle with respect
to the polar axis, three main wind components can be identified: a
hot, low density and extremely ionized flow in the polar region; a
dense, warm and fast equatorial outflow from the disk; and a
transition zone in which the disk outflow is hot and struggles to
escape the system.  The ionization state of the wind decreases from
polar to equatorial regions. Instead, the column densities increase
from polar to equatorial, up to very Compton-thick values
($N_H$$>$$10^{24}$~cm$^{-2}$).  The outflows can easily reach large
velocities, even higher than $\sim$$10^4$~km/s.

Lines of sight through the transition region of the simulated outflow,
where the density is moderately high
($n$$\sim$$10^8$--$10^{10}$~cm$^{-3}$) and the column density can
reach values up to $N_H$$\sim$$10^{24}$~cm$^{-2}$, result in spectra
that have considerable absorption features from ionized species
imprinted in the X-ray spectrum, mostly with intermediate/high
ionization parameters, log$\xi$$\sim$3--5~erg~s$^{-1}$~cm. This
strongly suggests that the absorption material could be observed in
the spectrum through Fe K-shell absorption lines from Fe XXV and Fe
XXVI (e.g., Sim et al.~2008; Schurch et al.~2009; Sim et al.~2010), in 
complete agreement with our detection of UFOs. 
In particular, Sim et al.~(2008) and Sim et al.~(2010) used their accretion disk wind model to successfully reproduce the 2--10~keV spectra of two bright radio-quiet AGN in which strong blue-shifted Fe K absorption lines were detected in their XMM-Newton spectra, namely Mrk~766 (from Miller et al.~2007) and PG~1211+143 (from Pounds et al.~2003 ). Notably, the authors have been able to account for both emission and absorption features in a physically self-consistent way and demonstrated that accretion disk winds/outflows might well imprint also other spectral signatures in the X-ray spectra of AGN (e.g., Pounds \& Reeves 2009 and references therein).

Hydrodynamic wind simulations are highly inhomogeneous in density,
column and ionization and have strong rotational velocity
components. Therefore the outflow, especially in its innermost
regions, is rather unstable.  In particular, the outflow properties
through the transition region show considerable variability and this
is expected to be reflected by the spectral features associated with
this region, i.e. by the corresponding blue-shifted Fe XXV/XXVI
K-shell absorption lines.

Proga \& Kallman (2004) and Schurch et al.~(2009) state that it is
possible that some parts or blobs of the flow, especially in the
innermost regions, do not have enough power to allow a ``true'' wind
to be generated. In these cases, a considerable amount of material is
driven to large-scale heights above the disk but the velocity of the
material is insufficient for it to escape the system and it will
eventually fall back onto the disk.  Despite returning to the
accretion disk at larger radii, while it is above the disk, this
material can imprint features on the observed X-ray spectrum
(e.g., Dadina et al.~2005 and references therein). This can indeed be
the case for some of the UFOs discussed here (i.e., 3C~111).

This overall picture is also partially in agreement with what
predicted by the ``aborted jet'' model by Ghisellini et
al.~(2004). This model was actually proposed to explain, at least in
part, the high-energy emission in radio-quiet quasars and Seyfert
galaxies.  It postulates that outflows and jets are produced by every
black hole accretion system. Blobs of material can then be ejected
intermittently and can sometimes only travel for a short radial
distance and eventually fall back, colliding with others approaching.
Therefore, the flow can manifest itself as erratic high-velocity
ejections of gas from the inner disk and it is expected that some
outflows/blobs are not fast enough to escape the system and will
eventually fall back onto the disk.  An intriguing possibility could
be that these outflows are generated by localized ejection of material
from the outer regions of a bubbling corona, which emits the bulk of
the X-ray radiation (Haardt \& Maraschi 1991), in analogy with what
observed in the solar corona during the Coronal Mass Ejection events
(e.g., Low 1996). The velocity and frequency of these strong events
should then be limited to some extent, in order not to cause the
disruption or evaporation of the corona itself.  Such extreme
phenomena could then be the signatures of the turbulent environment
close to the super-massive black hole.

The detection of UFOs in both radio-quiet and radio-loud galaxies
suggests a similarity of their central engines and demonstrates that
the presence of strong relativistic jets do not exclude the existence
of winds/outflows from the putative accretion disk. Moreover, it has
been demonstrated by Torresi et al.~(2010) and Reeves et al.~(2009)
that a warm absorber is indeed present also in BLRGs (in particular
3C~382) and this indicates that jets and slower winds/outflows can
coexist in the same source, even beyond the broad-line region.

However, BLRGs are radio-loud galaxies and they have powerful jets.
Therefore, the fact that for BLRGs we are observing down to the
outflowing stream at intermediate angles to the jet
($\sim$15--30$^{\circ}$; e.g.,  Eracleous \& Halpern 1998) suggests
that the fast winds/outflows we observe are at greater inclination
angles with respect to the jet axis, somewhat similar to what expected
for accretion disk winds (e.g., Proga \& Kallman 2004).  These outflows
would then not be able to undergo the processes which instead
accelerate the jet particles to velocities close to the speed of
light.

For instance, studies of Galactic stellar-mass black holes, or
micro-quasars, showed that wind formation occurs in competition with
jets, i.e. winds carry away matter halting their flow into jets
(e.g., Neilsen \& Lee 2009). Given the well-known analogy between
micro-quasars and their super-massive relatives, one would naively
expect a similar relationship for radio-loud AGN. The BLRGs 3C~111 and
3C~120 are regularly monitored in the radio and X-ray bands with the
VLBA and RXTE as part of a project aimed at studying the disk-jet
connection (e.g., Marscher et al~2002). We have detected UFOs in both of 
these sources (see Table~4), and indeed in both cases the 4--10~keV
fluxes measured with Suzaku corresponded to historical low(est) states
if compared to the RXTE long-term light curves. For instance,
correlated spectroscopic observations of 3C~111, where the shortest
variability timescales are predicted (t$\sim$7 days), during low and
high jet continuum states could provide, in a manner analogous to
micro-quasars, valuable information on the synergy among disk, jet,
and outflows, and go a long way towards elucidating the physics of
accretion/ejection in radio-loud AGN.

However, whether it is possible to accelerate such ultra-fast outflows
with velocities up to $\sim$0.15c only through UV line-driving is
unclear. Moreover, the material needs to be shielded from the high
X-ray ionizing flux in the inner regions of AGN, otherwise it would
become over-ionized and the efficiency of this process would be
drastically reduced.  Other mechanisms as well are capable of accelerating 
winds from accretion disks, in particular radiation pressure through Thomson
scattering and magnetic forces.

In fact, Ohsuga et al.~(2009) proposed a unified model of
inflow/outflow from accretion disks in AGN based on radiation-MHD
simulations. Disk outflows with helical magnetic fields, which are
driven either by radiation-pressure force or magnetic-pressure, are
ubiquitous in their simulations. In particular, in their case A (see
their Fig.~1) a geometrically thick, very luminous disk forms with a
luminosity $L \sim L_{Edd}$, which effectively drives a fast
Compton-thick wind with velocities up to $\sim$0.2--0.3c.  It is
important to note that the models of Ohsuga et al.~(2009) include both
radiation and magnetic forces which, depending on the state of the
system, can generate both relativistic jets and disk winds.

Moreover, King \& Pounds (2003) and King (2010) showed that black
holes accreting with modest Eddington rates are likely to produce fast
Compton-thick winds. They considered only radiation-pressure and
therefore fast winds can be effectively generated by low magnetized
accretion disks as well. In particular, King (2010) derived that
Eddington winds from AGN are likely to have velocities of $\sim$0.1c
and to be highly ionized, showing the presence of helium- or
hydrogen-like iron.  These properties strongly point toward an
association of our detected UFOs from the innermost regions of AGN
with Eddington winds/outflows from the putative accretion disk.

Depending on the estimated covering fraction, the derived mass outflow
rate of the UFOs can be comparable to the accretion rate and their
kinetic power can correspond to a significant fraction of the
bolometric luminosity and is comparable to the jet power.  Therefore,
the UFOs may have the possibility of bringing significant amount of
mass and energy outward, potentially contributing to the expected
feedback from the AGN.  In particular, King (2010) demonstrated that
fast outflows driven by black holes in AGN can explain important
connections between the SMBH and the host galaxy, such as the observed
$M_{BH}$--$\sigma$ relation (e.g., Ferrarese \& Merritt 2000).  These
UFOs can potentially provide an even more important contribution to the
expected feedback between the AGN and the host galaxy than the jets in
radio-loud sources. In fact, even if jets are highly energetic, they
are also extremely collimated and carry a negligible mass. Fast
winds/outflows from the accretion disks, instead, are found to be
massive and extend over wide angles. Thus, we suggest that UFOs in
radio-loud AGN are a new, important ingredient for feedback models
involving these sources.

\section{Summary and Conclusions}

Using high signal-to-noise Suzaku observations, we detected several 
absorption
lines in the $\sim$7--10~keV band of three out of five BLRGs with high 
statistical significance. If interpreted as blueshifted K-shell resonance 
absorption lines
from Fe XXV and Fe XXVI, the lines imply the presence of
outflowing gas from the central regions of BLRGs with mildly
relativistic velocities, in the range $\simeq$0.04--0.15c.  The inferred
 ionization states and column densities of the absorbers are in the range
log$\xi$$\sim$4--5.6~erg~s$^{-1}$~cm and
$N_H$$\sim$$10^{22}$--$10^{23}$~cm$^{-2}$, respectively.  This is the
first time that evidence for Ultra-fast Outflows (UFOs) from the
central regions of radio-loud AGN is obtained at X-rays. 

The estimated location of these UFOs at distances within 
$\sim$0.01--0.3~pc from the
central super-massive black hole suggests that the outflows might be
connected with AGN accretion disk winds/ejecta (e.g., King \& Pounds
2003; Proga \& Kallman 2004; Ohsuga et al.~2009; King 2010).
Depending on the covering fraction estimate (here, $C$$\sim$0.6),
their mass outflow rate can be comparable to the accretion rate and
their kinetic power may correspond to a significant fraction of the
bolometric luminosity and be comparable to the jet power.  These UFOs
would thus bring outward significant amounts of mass and energy,
potentially contributing to the expected feedback from the AGN on the
surrounding environment.

These results are in analogy with the recent findings of blue-shifted
Fe K absorption lines at $\sim$7--10~keV in the X-ray spectra of
several radio-quiet AGN, which demonstrated the presence of UFOs in
the central regions of these sources (e.g., APM~08279$+$5255, Chartas
et al.~2002; PG~1115$+$080, Chartas et al.~2003; PG~1211+143, Pounds
et al.~2003; IC4329A, Markowitz et al.~2006; MCG-5-23-16, Braito et
al.~2007; Mrk~509, Cappi et al.~2009; PDS~456, Reeves et al.~2009a;
see Tombesi et al.~2010 for a systematic study on a large sample of
Seyfert galaxies).  In particular, it is important to note that the
physical parameters of UFOs in radio-loud AGN previously discussed
are completely consistent with those reported in
radio-quiet AGN. This strongly suggests that we could actually be
witnessing the same physical phenomenon in the two classes of objects
and this can help us improve the understanding of the relation between
the disk and the formation of winds/jets in black hole accretion
systems.

Several questions remain open. It is important to note that the
estimate of the covering factor $C\sim0.6$ in \S~5.2 might actually be
only a lower limit.  Fast outflows are expected to
come from regions close to the central black hole and to be highly
ionized.  Thus, a slight increase in the ionization level of the
absorbers would cause iron to be completely ionized and the gas to
become invisible also in the Fe K band.  Therefore, it is also quite 
possible that most,
\emph{if not all}, radio-loud AGN contain ultra-fast outflows which
can not be seen at present simply because they are highly ionized.
 
The physical properties of UFOs in BLRGs are also of great interest to
understand the dynamics of accretion/ejection and the disk-jet
connection. 
In particular, by studying the source variability, which, in some sources, is 
expected to occur on timescales as short as a few days, we can investigate 
the gas densities and internal dynamics of the outflow, as well as 
better constrain 
its distance from the SMBH. This can help us understand in
detail whether the UFOs in radio-loud AGN are similar to those in
radio-quiet ones, or if major quantitative differences exist which
affect jet formation and thus the radio-loud/radio-quiet AGN division.

Finally, a substantial improvement is expected from the higher
effective area and supreme energy resolution (down to $\sim$2--5~eV)
in the Fe K band offered by the calorimeters on board the future
Astro-H and IXO missions. In particular, the lines will be resolved
and also their profiles could be measured. The parameters of UFOs will
be determined with unprecedented accuracy and their dynamics could, 
potentially, also be studied through time-resolved spectroscopy on short
time-scales (e.g., Tombesi et al.~2009).

\acknowledgments

We thank Laura Maraschi, Demos Kazanas, Keigo Fukumura, and Meg Urry
for useful discussions. FT and RMS acknowledge financial support from
NASA grant NAG5-10708 and the Suzaku program. MC acknowledge financial
support from ASI under contract I/088/06/0.

\appendix

\section{Notes on single sources}

\subsection{3C~111}

The 3.5--10.5~keV XIS-FI spectrum of 3C~111 is well described by a
simple power-law continuum (with $\Gamma\simeq1.5$) and a narrow
neutral Fe K$\alpha$ emission line at the rest-frame energy of 6.4~keV
(see Table~2). A detailed broad-band spectral analysis of the Suzaku
spectrum of this source will be reported in Ballo et al.~(2010, in
prep.).  However, as it can be seen from the ratio of the spectrum
against a simple power-law continuum reported in the upper panel of
Fig.~1 (left), further complexities are present in the spectrum. In
fact, besides the narrow emission line, two absorption features can be
clearly seen at the observed energies of $\sim$7~keV and
$\sim$8--9~keV.  These absorption features are still present in the
energy-intensity contour plot (see lower panel of Fig.~1, left), which
suggest that their detection confidence levels should be higher than
99\%.

Therefore, we directly fit the data, adding
two further absorption lines to the baseline model.  The detailed line
parameters are reported in Table~3.  The first absorption line is not
resolved and is detected at a rest-frame energy of
E$=$$7.26\pm0.03$~keV, with an equivalent width of
EW$=$$-31\pm15$~eV. Its detection confidence level is high: 99.9\%
from the standard F-test and 99\% from extensive Monte Carlo
simulations (see \S~3.3).  The most intense spectral features
expected at energies $\ga$7~keV are the inner K-shell resonances from
Fe XXVI (e.g., Kallman et al.~2004). These lines are those of the Lyman
series, that is: the Ly$\alpha$ (1s--2p) at E$=6.966$~keV, the
Ly$\beta$ (1s--3p) at E$=8.250$~keV, the Ly$\gamma$ (1s--4p) at
E$=8.700$~keV and the Ly$\delta$ (1s--5p) at E$=8.909$~keV (all line
parameters have been taken from the
NIST\footnote{http://physics.nist.gov/PhysRefData/ASD/lines\_form.html}
atomic database, unless otherwise stated).  However, the observed line energy
is not consistent with any of these known atomic transitions.  If
identified with Fe XXVI Ly$\alpha$ resonant absorption, the centroid
of the line indicates a substantial blue-shifted velocity of
$+0.041\pm0.003$c.

The second absorption line is at a measured rest-frame energy of
E$=$$8.69^{+0.13}_{-0.08}$~keV. It is broader than the first one, with
a resolved width of $\sigma$$=$$390^{+270}_{-70}$~eV and an
equivalent width of EW$=$$-154\pm80$~eV (see Table~3). The detection confidence
level of the line is higher than 99.9\% with both the F-test and Monte
Carlo simulations (see \S3.3).  Also in this case the energy of the
line is not consistent with any known atomic transition. If identified
with Fe XXVI Ly$\beta$, the centroid of the line indicates a
blue-shifted velocity of $\sim$0.05c. This value is comparable with
that of the former line.  However, if this is the case, the ratio of
the EWs of the Fe XXVI Ly$\alpha$ and Ly$\beta$ would be
$\sim$0.2. This is at odds with what expected from theory. In fact, the
ratio between these lines must be instead equal to $\simeq$5 (which is
the ratio of their oscillator strengths: $0.42$ and $0.08$,
respectively) and it could decrease to a minimum of $\simeq$1 when the
lines are substantially saturated (e.g., Risaliti et al.~2005).  This
would suggest that the second broad absorption line could actually be
a blend of different blue-shifted resonance lines, such as the
Ly$\beta$, Ly$\gamma$ and Ly$\delta$.  This scenario is supported by
the fact that the energy resolution of the XIS instruments degrades
with increasing energy (at E$\sim$8--9~keV it is of the order of
FWHM$\ga$200~eV) and therefore these lines could not be separated
properly.

To test whether a line blend is consistent with the data, we performed 
a fit adding to the baseline
model four additional narrow absorption lines with energies fixed to
the expected values for the Fe XXVI Lyman series and leaving their
common energy shift as a free parameter. These lines provide a very
good modeling of both absorption features at E$>$7~keV, with a global
$\Delta\chi^2=42$ for five additional parameters. The
probability of having these four absorption lines at these exact
energies simply from random fluctuations is very low, about $10^{-8}$.
Interestingly enough, their common blue-shifted velocity is
$+0.041\pm0.004$c, consistent with the one calculated above for the
first absorption line.  The resultant EWs of the four Fe XXVI
lines are: EW$=-25\pm8$~eV for the Ly$\alpha$, EW$=-35\pm14$~eV
for the Ly$\beta$, EW$=-27\pm-16$ for the Ly$\gamma$ and EW$>-60$~eV
for the Ly$\delta$.  Their ratios are now consistent with the
theoretical expectations and the fact that they are close to unity suggests 
possible saturation effects.

In order to have a more physically consistent modeling of these
spectral features, we performed a fit using the Xstar photo-ionization
grid discussed in \S~3.4.  The best fit parameters are reported in
Table~4.  We obtained a good fit with a highly ionized absorber
($\Delta\chi^2$$=$22 for three additional parameters, required at a level of 
$>$99.9\%) with an
ionization parameter of log$\xi=5.0\pm0.3$~erg~s$^{-1}$~cm and a
column density of $N_H>2\times 10^{23}$~cm$^{-2}$.  The blue-shifted
velocity is $+0.041\pm0.003$c, completely consistent with the value
determined above fitting with four simple inverted Gaussian lines.
Given the extremely high ionization level of this absorbing material,
no other signatures are expected at lower energies as all the elements
lighter than iron are fully ionized, as indeed observed.

We conclude that the detected absorption features are actually due
to blue-shifted Fe XXVI Lyman series lines. In Fig.~2 (left) we plot the 
baseline model composed by a power-law continuum and a neutral 
Fe K$\alpha$ emission line, and
superimposed on it the model of the absorption features with two simple
inverted Gaussian lines and the Xstar model.  The plot shows that the
two models are almost completely coincident up to the first absorption
line (identified as Fe XXVI Ly$\alpha$ at the observed energy of
$\sim$7~keV) and clearly demonstrates that the apparent broadening of
the second absorption feature is actually due to a blend of the
three higher order Lyman series lines (i.e. Ly$\beta$, Ly$\gamma$ and
Ly$\delta$).

A consistency check of the Fe K absorber parameters from a broad-band XIS$+$PIN fit is reported in \S4.2. Instead, the XIS-FI background analysis and the consistency check of the line parameters among the different XIS cameras (see Table~5), along with the discussion of possible alternative modelings of the lines, are reported in Appendix B.

\subsection{3C~390.3}

A power-law continuum (with $\Gamma\simeq1.6$) plus a narrow neutral
Fe K$\alpha$ emission line at 6.4~keV provide a good modeling of the
3.5--10.5~keV XIS-FI spectrum of 3C~390.3 (see Table 2). The
broad-band spectral analysis of this Suzaku data set has been reported
by Sambruna et al.~(2009).
From the spectral ratios and the energy-intensity contour
plots of Fig.~1 (right) there is indication of a possible narrow
absorption feature at the observed energy of $\sim$7.7~keV, with a
detection confidence level greater than 99\%.

Therefore, we fit the spectrum adding a
further narrow (unresolved) absorption line to the baseline model. The
rest-frame energy of the line is E$=$$8.11\pm0.07$~keV and its
equivalent width is EW$=$$-32\pm16$~eV (see Table 3). The detection
confidence level of the line is high: 99.9\% from the standard F-test
and 99.5\% from extensive Monte Carlo simulations (see \S~3.3).  Also
in this case the energy of the line is not consistent with any known
atomic transition. However, the most intense lines expected from a
highly ionized absorber at E$\ga$7~keV are the Fe XXVI Lyman series
(see \S~A.1.).  If identified with Fe XXVI Ly$\alpha$ resonant
absorption, the centroid of the line indicates a substantial
blue-shifted velocity of $+0.150\pm0.005$c.  In order to derive a more
physically consistent modeling of this absorption line we performed a
fit using the Xstar photo-ionization grid discussed in \S3.4.  The
best fit parameters are reported in Table~4.  We obtained a good fit
with a highly ionized absorber ($\Delta\chi^2$$=$14 for three more parameters,
 required at the $\simeq$99.5\% level) with an ionization parameter of
log$\xi=5.6^{+0.2}_{-0.8}$~erg~s$^{-1}$~cm and a column density of
$N_H>3\times 10^{22}$~cm$^{-2}$. The blue-shifted velocity of the
absorber is $+0.146\pm0.004$c, completely consistent with what was
derived from fitting with a simple inverted Gaussian line.  Given
the very high ionization level of this absorbing material, no other
significant signatures are expected at lower energies as all the
elements lighter than iron are completely ionized.

The comparison of the best fit results for 3C~390.3 including the
baseline model and superimposed the modeling of the blue-shifted
absorption line (identified as Fe XXVI Ly$\alpha$) with a simple
narrow inverted Gaussian or with the Xstar photo-ionization code is
shown in Fig.~2 (right). The two models coincide completely, apart
from a few weak higher order Lyman series resonances which cannot be
detected with sufficient significance given the quality of the
spectral data.

In \S4.2 we report a consistency check of the results performing also a 
broad-band XIS$+$PIN spectral analysis. In Appendix B we discuss the 
XIS-FI background analysis, the consistency check of the blue-shifted Fe K 
absorption lines among the different XIS cameras (see Table~5) and also 
possible alternative modelings.

\subsection{3C~120}

The 3.5--10.5~keV XIS-FI spectrum of observation 3C~120a (see \S~2) is well
modeled by a power-law continuum ($\Gamma\simeq 1.75$) and a narrow
neutral Fe K$\alpha$ emission line at the rest-frame energy of 6.4~keV
(see Table~2).  As it can be seen from the ratio of the spectrum
against a power-law continuum and the contour plots in the left part
of Fig.~3 (upper and lower panels, respectively) there are no
significant emission/absorption features in the Fe K band apart from
the narrow Fe K$\alpha$ emission line.

The 3.5--10.5~keV XIS-FI spectrum of observation 3C~120b (see \S~2) is well
described by a power-law continuum (with $\Gamma\simeq 1.6$) plus a
narrow neutral Fe K$\alpha$ emission line at E$\simeq$6.4~keV and a further 
narrow emission line at E$\simeq$6.9~keV (see Table~2). These overall
results are in agreement with the spectral analysis of this data set
previously reported by Kataoka et al.~(2007).  However, the spectral
ratio and the energy-intensity contour plots in the right part of
Fig.~3 (upper and lower panels, respectively), suggest that further
complexities might be present in the Fe K band.  In particular, 
there is evidence for absorption structures at the observed energies of 
$\sim$7--7.4~keV and $\sim$8--9~keV. The
contours in the right part of Fig.~3 (lower panel) suggest that their 
detection confidence levels are higher than 99\%.

A direct spectral fitting revealed that the absorption
structures at $\sim$7--7.4~keV are actually composed of two narrow
(unresolved, $\sigma=$10~eV) absorption lines.  They are detected at
rest-frame energies of E$=$$7.25\pm0.03$~keV and
E$=$$7.54\pm0.04$~keV, respectively. Their equivalent widths are
EW$=$$-10\pm5$~eV and EW$=$$-12\pm6$~eV, respectively. Their detection
confidence level is $\simeq$99\% from the F-test, which slightly
reduces to 91\% and 92\% from Monte Carlo simulations (see \S3.3). The 
detailed line parameters are listed in Table~3.
Their energies are not consistent with any known atomic transition.
However, their location in the spectrum and their energy spacing
suggest a possible identification with blue-shifted resonance
absorption lines from Fe XXV He$\alpha$ (1s$^2$--1s2p) at E$=$6.697~keV
and Fe XXVI Ly$\alpha$ (1s--2p) at E$=$6.966~keV.  Their corresponding
blue-shifted velocities are substantial and consistent one with each
other, i.e. $+0.076\pm0.003$c and $+0.076\pm0.004$c, respectively.  

The second absorption structure that is observed at the energy of
E$\sim$8--9~keV is broad. If modeled with a simple inverted Gaussian,
the resultant rest-frame energy is E$=$$8.76\pm0.12$~keV, with a
broadening of $\sigma$$=$$360^{+160}_{-120}$~eV and equivalent width of
EW$=$$-50\pm13$~eV. Its detection confidence level is 99.9\% from the F-test 
and slightly reduces to 99.8\% with Monte Carlo simulations (see Table~3).  
Also in this
case the energy of the line is not consistent with any known atomic
transition.  However, from the identification of two previous
absorption lines, we can infer the possible presence of other
resonance features from the same ionic species.  In fact, the lower
energy resolution of the instrument at those energies
(FWHM$\ga$200~eV) and the spacing with respect to the first two lines
suggest this broad absorption structure could actually be a blend
of at least two further narrow resonance lines, namely Fe XXV He$\beta$
(1s$^2$--1s3p) at E$=$7.88~keV and Fe XXVI Ly$\beta$ (1s--3p) at
E$=$8.25~keV.

To test the consistency of this global line identification, we
performed a fit adding to the baseline model four narrow absorption
lines with energies fixed to the expected values for these Fe XXV and
Fe XXVI resonances and leaving their common energy shift as a free
parameter.  This provided a very good modeling of all the absorption
structures at E$\ga$7~keV, with a $\chi^2$ improvement of 25 (for five
additional parameters). The global probability to have these four
absorption lines at these exact energies simply from random
fluctuations is low, about $4\times10^{-4}$. Interestingly, their
common blue-shifted velocity is $+0.076\pm0.003$, completely consistent
with what derived fitting each line separately. The resultant EWs of
these lines are: EW$=$$-10\pm5$~eV for the Fe XXV He$\alpha$,
EW$=$$-11\pm8$~eV for the Fe XXV He$\beta$, EW$=$$-11\pm7$~eV for the
Fe XXVI Ly$\alpha$ and EW$=$$-13\pm9$~eV for the Fe XXVI
Ly$\beta$. Their relative ratios are of the order of unity, which
would suggest possible saturation effects.

Finally, we performed a fit using the Xstar photo-ionization grid
discussed in \S3.2 in order to have a more physically consistent
modeling of these spectral features. The best fit parameters are
reported in Table~4.  We obtained a good fit with a highly ionized
absorber ($\Delta\chi^2$$=$12 for three more parameters, required at a level 
of $\simeq$99\%) with
an ionization parameter of log$\xi$$=$$3.8\pm0.2$~erg~s$^{-1}$~cm and
a total column density of $N_H$$=$$1.1^{+0.5}_{-0.4} \times
10^{22}$~cm$^{-2}$.  This model simultaneously takes into account all
the four absorption features we discussed previously.  The resultant
blue-shifted velocity is $+0.076\pm0.003$c, completely consistent with the value
 estimated by fitting the lines with simple inverted Gaussians.
Given the high ionization level of this absorbing material, no other
significant signatures are expected at lower energies as all the
elements lighter than iron are almost completely ionized.

The conclusion that the detected absorption features are actually due
to Fe XXV and Fe XXVI resonant lines is well represented in
Fig.~4. Here we can see the best fit baseline model composed of a
power-law continuum and a neutral Fe K$\alpha$ emission line and the
superimposed modeling of the absorption structures with two narrow and
one broad inverted Gaussians or with the physically self-consistent
Xstar photo-ionization code.  The plot shows that the two models are
completely coincident up to the first two narrow absorption lines
(identified as blue-shifted Fe XXV He$\alpha$ and Fe XXVI Ly$\alpha$)
and clearly demonstrates that the broad absorption structure at the higher
energy is actually composed by several narrow resonant lines from the
same ionic species (i.e. mainly Fe XXV He$\beta$ and Fe XXVI Ly$\beta$)
which appear to be blended together due to the lower instrumental
resolution and signal-to-noise in this energy band (this is similar to
the conclusion drawn for 3C~111 in \S~A.1).

We also checked for variability of the blue-shifted Fe K absorption lines
between observations 3C~120a and 3C~120b.  We added three
absorption lines to the base line model of observation 3C~120a, with
energies and widths fixed to those of observation 3C~120b, and calculated the
90\% lower limits on the equivalent widths.  The values are reported
in Table~3. Unfortunately, the lower S/N in
observation 3C~120a alone does not allow us to affirm that the lack of
absorption lines in this observation was due to temporal variability.

The consistency of the results from a broad-band XIS$+$PIN spectral analysis is presented in \S4.2. Instead, in Appendix B we discuss the XIS-FI instrumental background, the consistency of the blue-shifted absorption lines parameters among the different XIS cameras (see Table 5) and their possible alternative identifications.

\subsection{3C~382}

The 3.5--10.5~keV XIS-FI spectrum of 3C~382 is well represented by a
power-law continuum (with $\Gamma\simeq1.75$) plus a narrow neutral Fe
K$\alpha$ emission line at E$=$6.4~keV and a further weak narrow emission 
line at E$\simeq$6.9~keV (see Table~2). A broad-band spectral analysis of this 
Suzaku data set will
be reported in Ballo et al.~(2010, in prep.).  From the spectral ratio
and the energy-intensity contour plots of Fig.~5 (left panel) it can
be seen that an additional narrow weak emission line at the rest-frame
energy of $\sim$7~keV is observable (we refer the reader to Ballo et
al.~2010, in prep.). However, there are no significant absorption
structures at energies $\ga$7~keV. We estimated the lower limit for
the presence of a narrow blue-shifted absorption line at the
indicative energy of 8~keV to be EW$>-20$~eV (see Table~3).

\subsection{3C~445}

The 3.5--10.5~keV XIS-FI spectrum of 3C~445 is affected by
substantial absorption by neutral/mildly-ionized material intrinsic to the
AGN. The baseline model is composed by a power-law continuum (with
$\Gamma\simeq1.6$) absorbed by neutral material ($N_H\simeq2\times
10^{23}$~cm$^{-2}$) and a narrow neutral Fe K$\alpha$ emission line at
6.4~keV (see Table 2). The broad-band spectral analysis of this Suzaku
data set will be reported in Braito et al.~(2010, in prep.).  From the
spectral ratios and the energy-intensity contour plots of Fig.~5
(right panel) there is indication for a possible narrow weak emission
feature red-ward to the Fe K$\alpha$ line (we refer the reader to
Braito et al.~2010, in prep.). We estimated the lower limit for
the presence of a narrow blue-shifted absorption line at the
indicative energy of 8~keV to be EW$>-45$~eV (see Table 3).

\section{XIS-FI background and consistency checks}

The background level for these bright sources in the 7--10~keV band is 
negligible, always less than 10\% of the source counts (see Table~1). 
However, it is important to note that the XIS cameras have a few
instrumental background emission lines at energies
greater than 7~keV, the most intense of which is the Ni K$\alpha$ at 
E$=$7.47~keV (Yamaguchi et al.~2006).  
These lines originate from the interaction of the cosmic rays with the sensor 
housing and electronics and this also causes their intensity to be slightly 
dependent on the location on the detector. 
Therefore, the selection of the background on a region of the CCD
where the intensity of the lines is slightly higher/lower than that of
those actually on the source extraction region can possibly induce
spurious absorption/emission lines in the background subtracted
spectrum. We performed some tests to check this possibility. 

First, since the XIS background emission lines are present at specific
energies (see Table~1 of Yamaguchi et al.~2006), we checked that the
observed energies of the absorption lines are indeed not consistent
with those values (see Table~3).  If the background is not subtracted
from the source spectrum, these lines would show up as weak emission
lines. Secondly, we then checked that the intensity of the emission
lines in the background and in the source spectrum without background
subtraction are indeed consistent. Thirdly, and finally, we inspected
that the values of the energy and equivalent width of the absorption
lines at E$>$7~keV in the source spectrum (see Table~3) are consistent
(within the 1$\sigma$ errors) with or without background subtraction.
These tests assure that our results on the absorption lines detected
in the 7--10~keV band are indeed not affected by any contamination
from the XIS instrumental background.

In Table~5 we report a consistency check of the absorption lines detected in 3C~111, 3C~390.3 and 3C~120b among the different XIS instruments. 
The values have been derived by independently fitting the XIS~0, XIS~2 (when available), XIS~3 and XIS-BI spectra. 
The lower S/N of the separate XIS spectra does not allow to clearly detect the absorption lines in each spectrum. However, the parameters are always consistent with those reported in Table~3.
Moreover, it is worth mentioning that in 3C~111 the blue-shifted Fe K absorption lines are detected independently in both of the XIS-FI CCDs (i.e. XIS 0 and 3), as shown in Table~5.  
This demonstrates that the line parameters derived from the different instruments are indeed consistent one with each other and assures the absence of any systematics induced by the combination of the XIS-FI spectra.

The search for narrow absorption lines in the $\sim$7--10~keV energy band could be complicated by the presence of ionized Fe K edges at energies in the range from E$\simeq$7.1~keV to E$\simeq$9.3~keV, depending on the ionization state of iron (from neutral to H-like). 
Hence, one could object that some of the spectral structures we identified as blue-shifted absorption lines could actually be interpreted equally as well by ionized Fe K edges. 
As a sanity check, we tested that the alternative modelling of the Gaussian absorption lines with simple sharp absorption edges (\emph{zedge} in XSPEC) did not significantly improve the spectral fits, as expected from the narrowness of the observed spectral features. 
Moreover, it is important to note that the commonly held view of sharp Fe K edges is an oversimplification of the real process and could lead to misleading results. 
In fact, it has been demonstrated that if the adequate treatment of the decay pathways of resonances converging to the K threshold is properly taken into account, the resulting edges are not sharp but smeared and broadened (e.g., Palmeri et al.~2002; Kallman et al.~2004). This effect can be negligible for neutral or extremely ionized iron (He/H-like) but is quite relevant for intermediate states (with energies in the range E$\simeq$7.2--9~keV). 
Intense Fe K resonance absorption lines from different ionization states would be expected to accompany the edges. 
Moreover, a proper characterization of the possible Fe K edges has already been taken into account when modeling the absorption features with the photo-ionization code Xstar.

Finally, it is worth noting that even if the cosmic abundance of nickel is
negligible with respect to that of iron ($\sim$5\%, from Grevesse et
al.~1996), the K-shell transitions of this element are distributed at
energies greater than 7~keV and could, in principle, complicate our
line identification.  However, contamination by  mildly
ionized Ni K$\alpha$ lines is very unlikely, as it would require
extremely high column densities
($N_H$$>$$10^{24}$--$10^{25}$~cm$^{-2}$) for these lines to be intense
enough to be observable, which would consequently generate strong
absorption lines and edges from all the other lighter elements as
well.  

The only possible contamination could be due to He/H-like Ni, whose 1s--2p transitions are at rest-frame energies of E$\simeq$7.8~keV and E$\simeq$8.1~keV, respectively. 
Also in this case the column densities required to have lines with measurable intensities would be extremely high ($N_H$$>$$10^{24}$--$10^{25}$~cm$^{-2}$).
However, the very high ionization level required to have significant columns of these ions are so extreme (log$\xi$$\ga$6~erg~s$^{-1}$~cm) that all the lighter elements would be completely ionized, with iron being the only possible exception, and therefore they will not contribute with other absorption features. 
We found that only the absorption line detected in 3C~390.3 at the energy of E$\simeq$8.11~keV could be associated with rest-frame absorption from H-like Ni (see Table~3). 
If this unlikely identification is correct, it would indicate the presence of an extremely Compton-thick, extremely ionized and static absorber in the central regions of this BLRG.
However, we state that the consistency of the line energy with H-like Ni is most probably only a mere coincidence. This is strengthened by the fact that none of the other lines detected at E$>$7~keV have energies consistent with those from highly ionized nickel.  
The same conclusion has been reached also by Tombesi et al.~(2010) who performed a systematic search for blue-shifted Fe K absorption lines in a large sample of radio-quiet AGN observed with XMM-Newton.

\clearpage

   \begin{figure*}[t]
   \centering
    \includegraphics[width=6cm,height=7.5cm,angle=0]{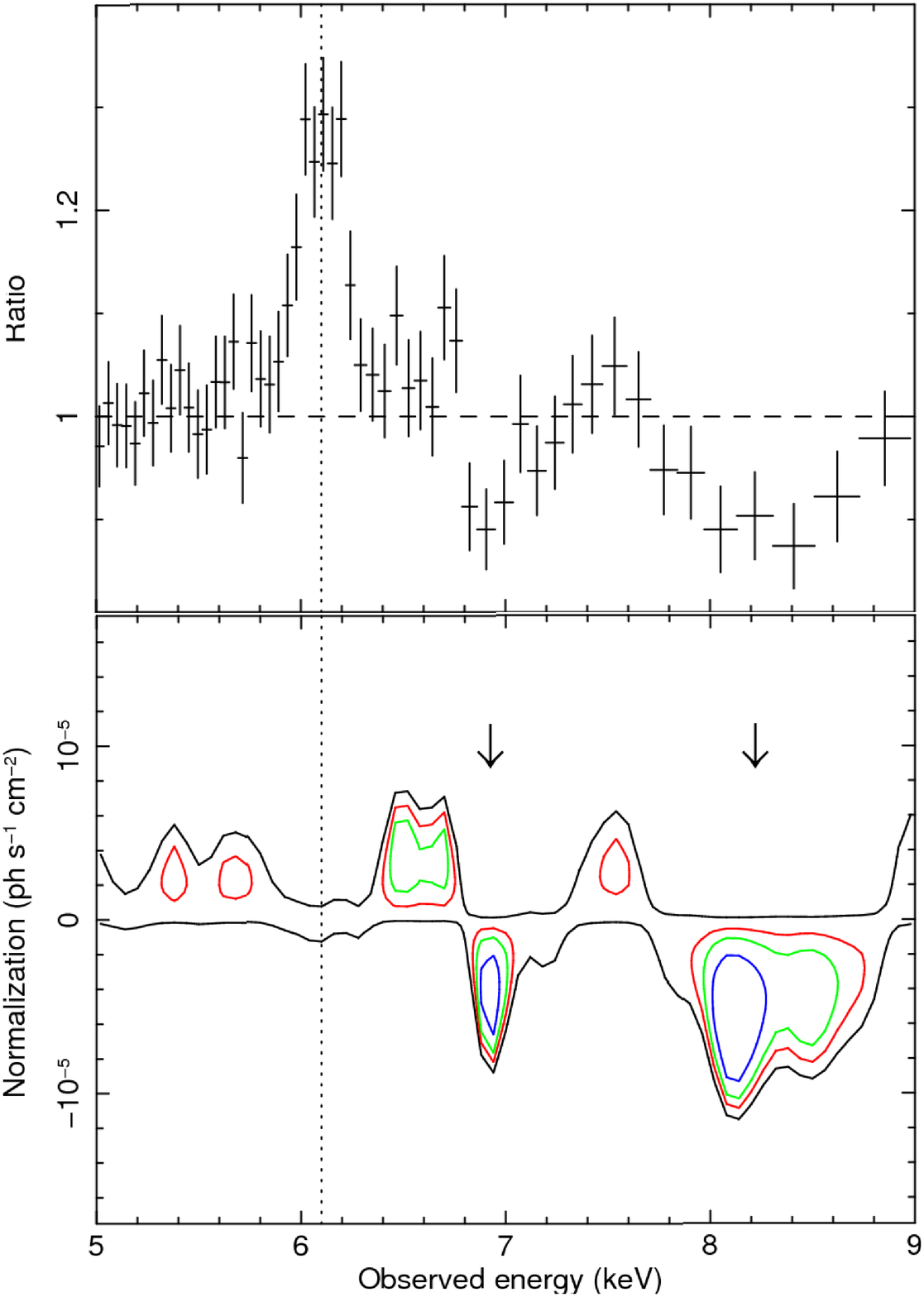}
\hspace{0.5cm}
   \includegraphics[width=6cm,height=7.5cm,angle=0]{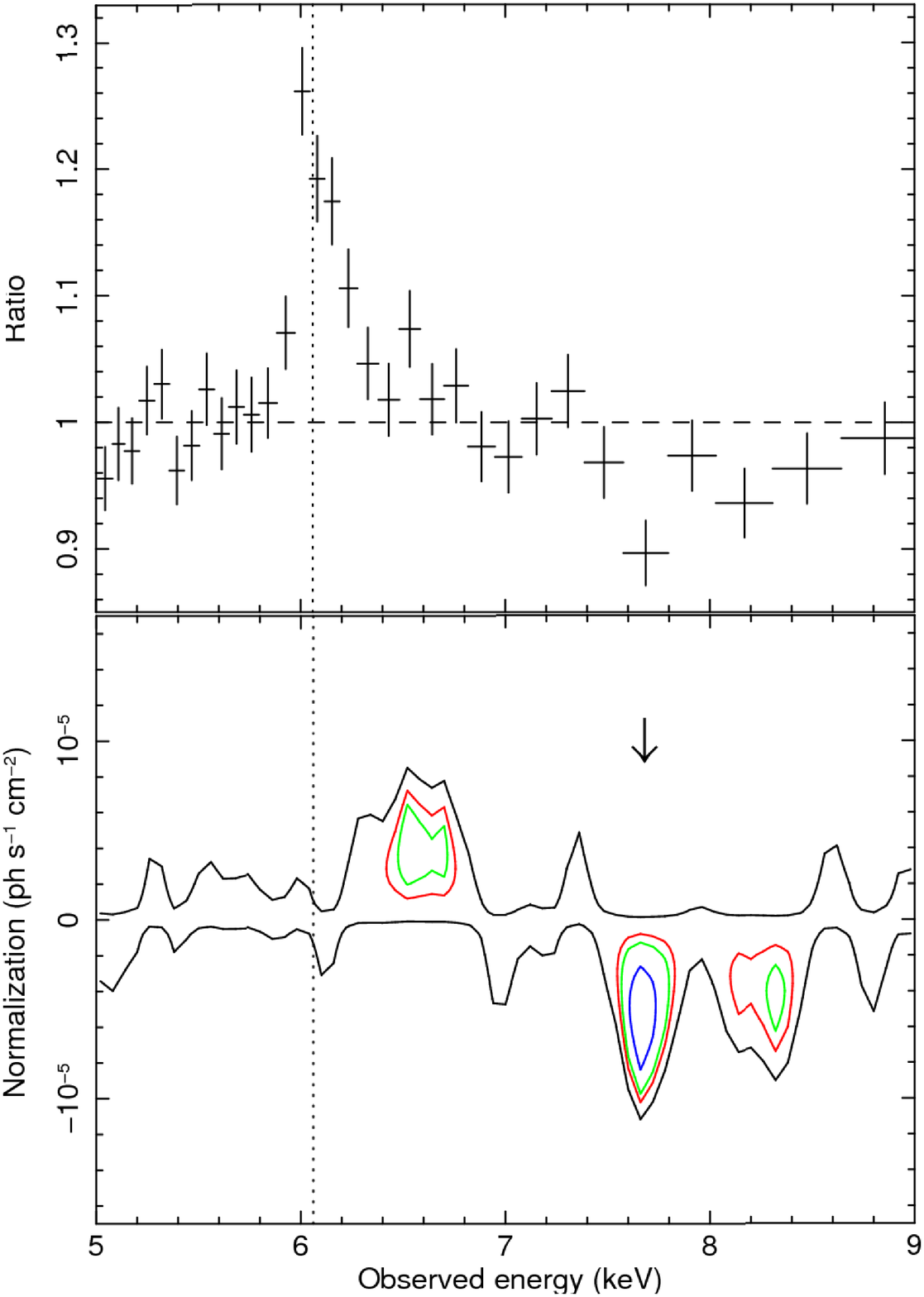}
   \caption{Suzaku XIS-FI spectra of 3C~111 (\emph{left}) and 3C~390.3 (\emph{right}) zoomed in the 5--9~keV band to emphasize the Fe K complex. \emph{Upper panel:} ratio against a power-law continuum. For plotting purposes only, data have been further grouped with XSPEC to reach a S/N in each energy bin of 21 and 32 for 3C~111 and 3C~390.3, respectively. \emph{Lower panel:} energy-intensity contours with respect to the baseline model described in Table~2 (see \S~3.1 and \S3.2 for more details), the arrows indicate the location of the blue-shifted absorption features (see A.1. and A.2. for more details).}
    \end{figure*}

   \begin{figure*}[ht]
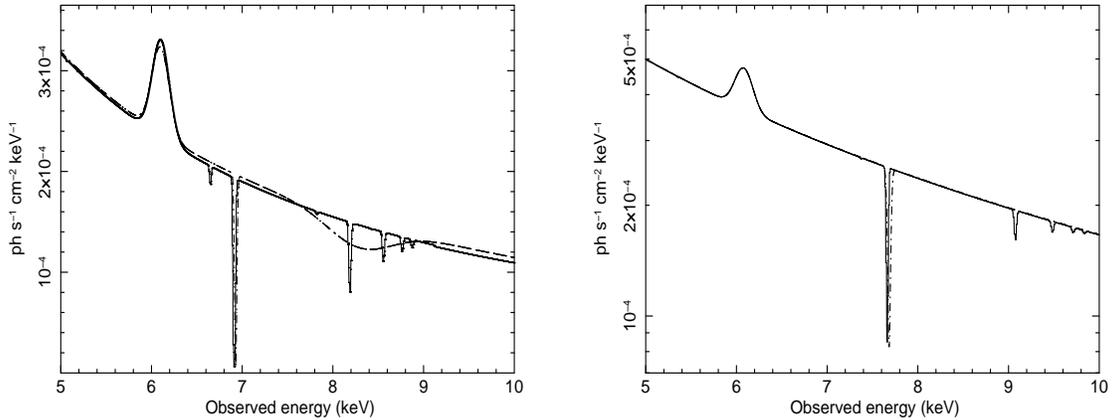

   \centering
    \includegraphics[width=5.5cm,height=7cm,angle=270]{figure_2l_bh.ps}
\hspace{0.5cm}
    \includegraphics[width=5.5cm,height=7cm,angle=270]{figure_2r_bh.ps}
   \caption{Comparison of the best fit model for 3C~111 (\emph{left}) and 3C~390.3 (\emph{right}) including the baseline model (see Table~2) plus the Gaussian absorption lines listed in Table~3 (dashed line) or the detailed photo-ionization modeling of the absorber reported in Table~4 (solid line).}
    \end{figure*}

   \begin{figure*}[ht]
   \centering
     \includegraphics[width=6cm,height=7.5cm,angle=0]{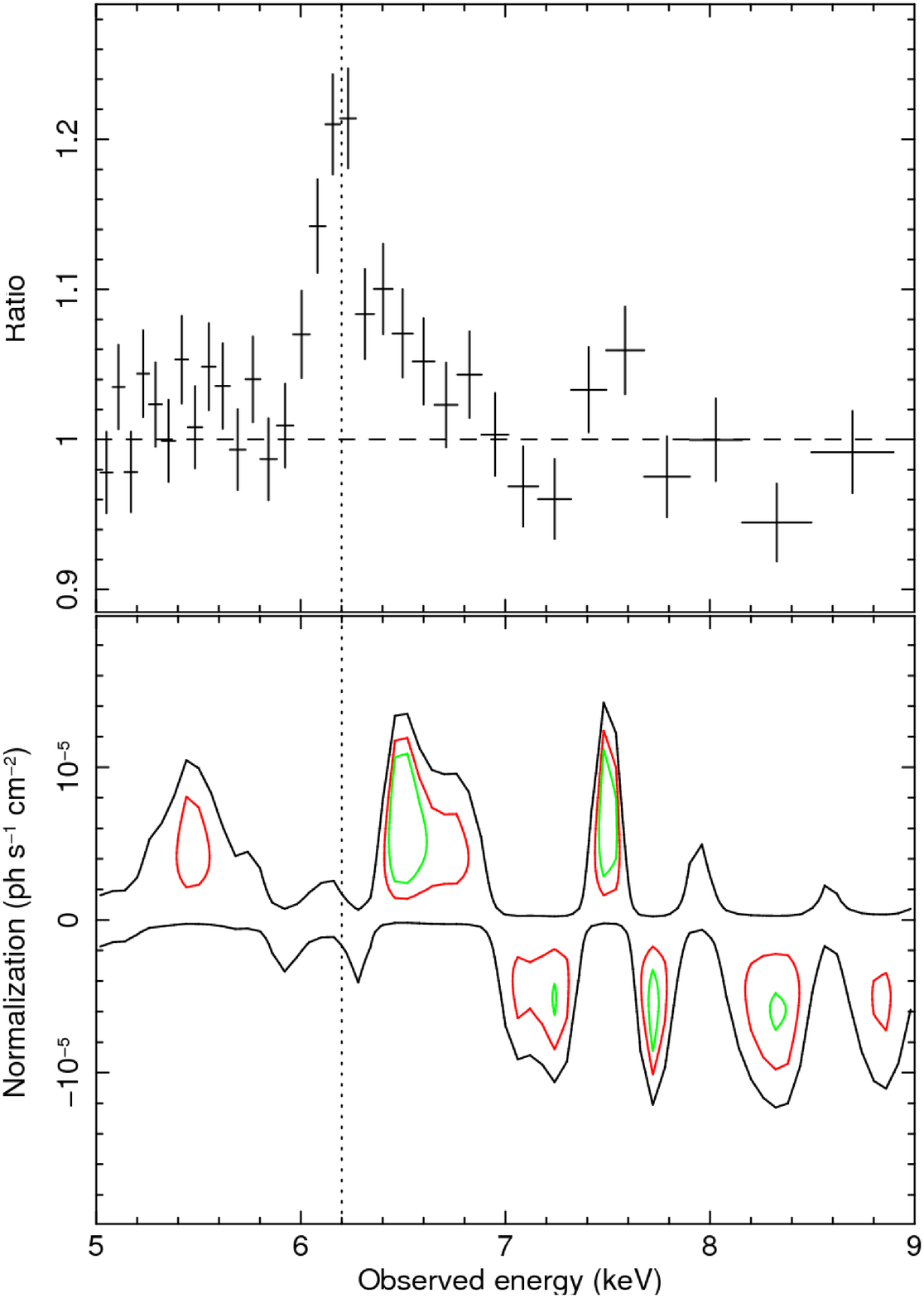}
\hspace{0.5cm}
   \includegraphics[width=6cm,height=7.5cm,angle=0]{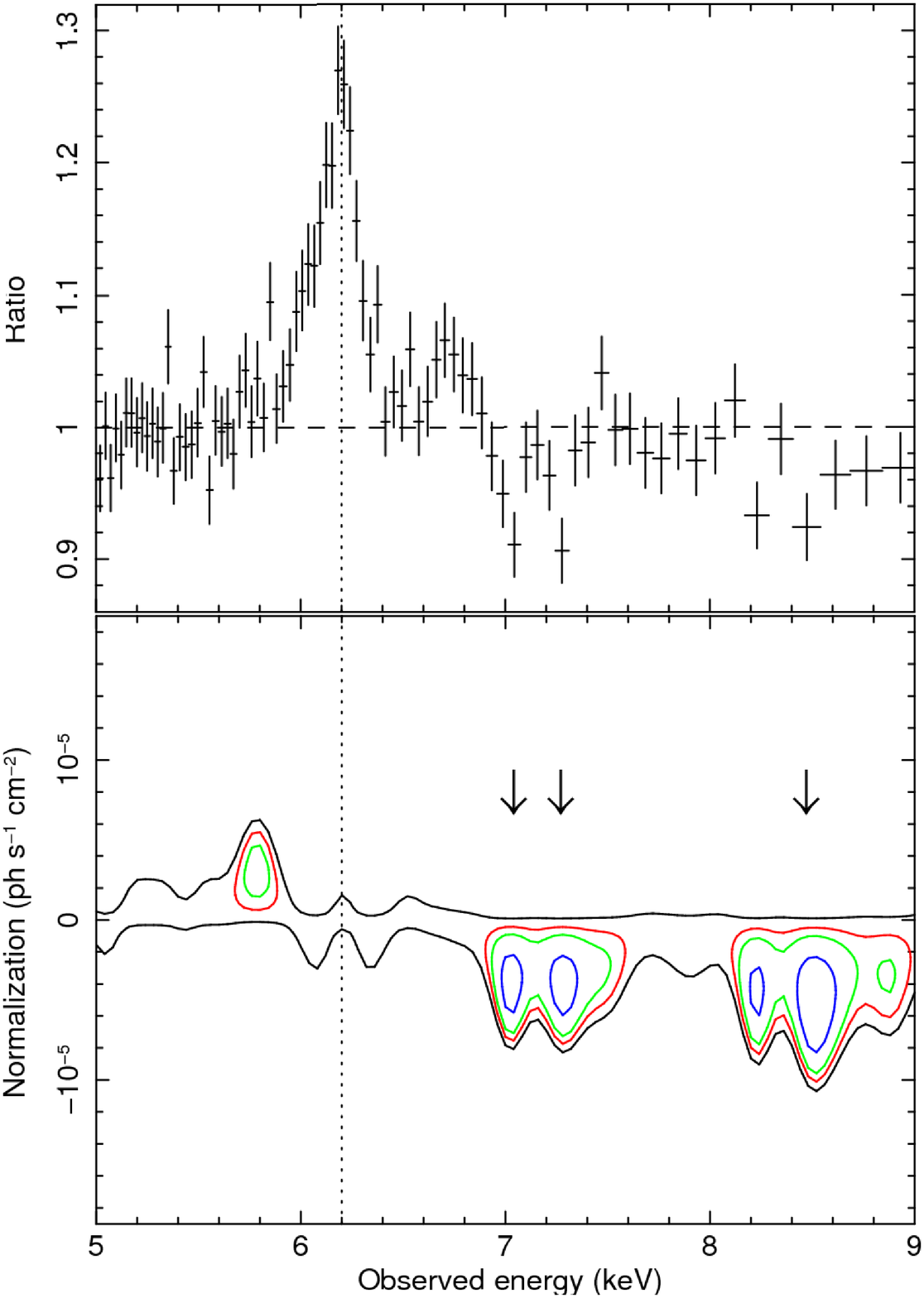}
   \caption{Suzaku XIS-FI spectra of 3C~120 (observations 3C~120a and 3C~120b on the \emph{left} and \emph{right}, respectively) zoomed in the 5--9~keV band to emphasize the Fe K complex. \emph{Upper panel:} ratio against a power-law continuum. For plotting purposes only, data have been further grouped with XSPEC to reach a S/N in each energy bin of 36 and 38 for observation 3C~120a and 3C~120b, respectively. \emph{Lower panel:} energy-intensity contours with respect to the baseline model described in Table~2 (see \S~3.1 and \S~3.2 for more details), the arrows indicate the location of the blue-shifted absorption features (see A.3. for more details).}
    \end{figure*}

   \begin{figure}[ht]
   \centering
    \includegraphics[width=5.5cm,height=7cm,angle=270]{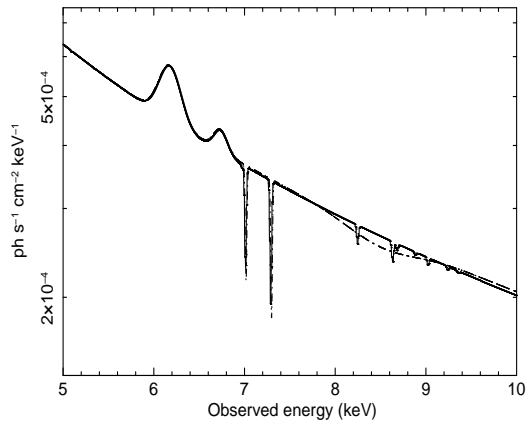}
   \caption{Comparison of the best fit model for 3C~120b including the baseline model (see Table~2) plus the Gaussian absorption lines listed in Table~3 (dashed line) or the detailed photo-ionization modeling of the absorber reported in Table~4 (solid line).}
     \end{figure}

   \begin{figure*}[ht]
   \centering
    \includegraphics[width=6cm,height=7.5cm,angle=0]{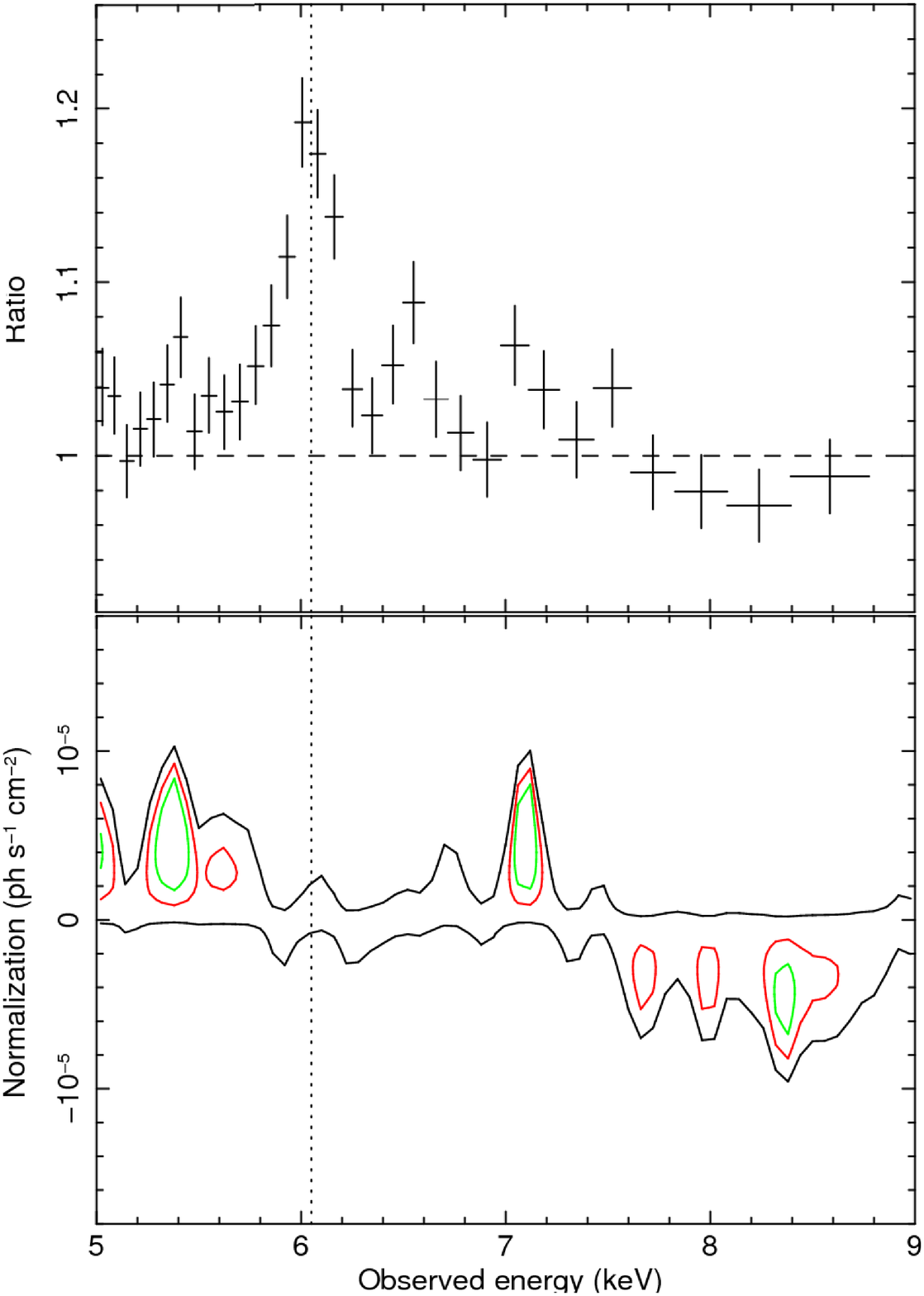}
\hspace{0.5cm}
   \includegraphics[width=6cm,height=7.5cm,angle=0]{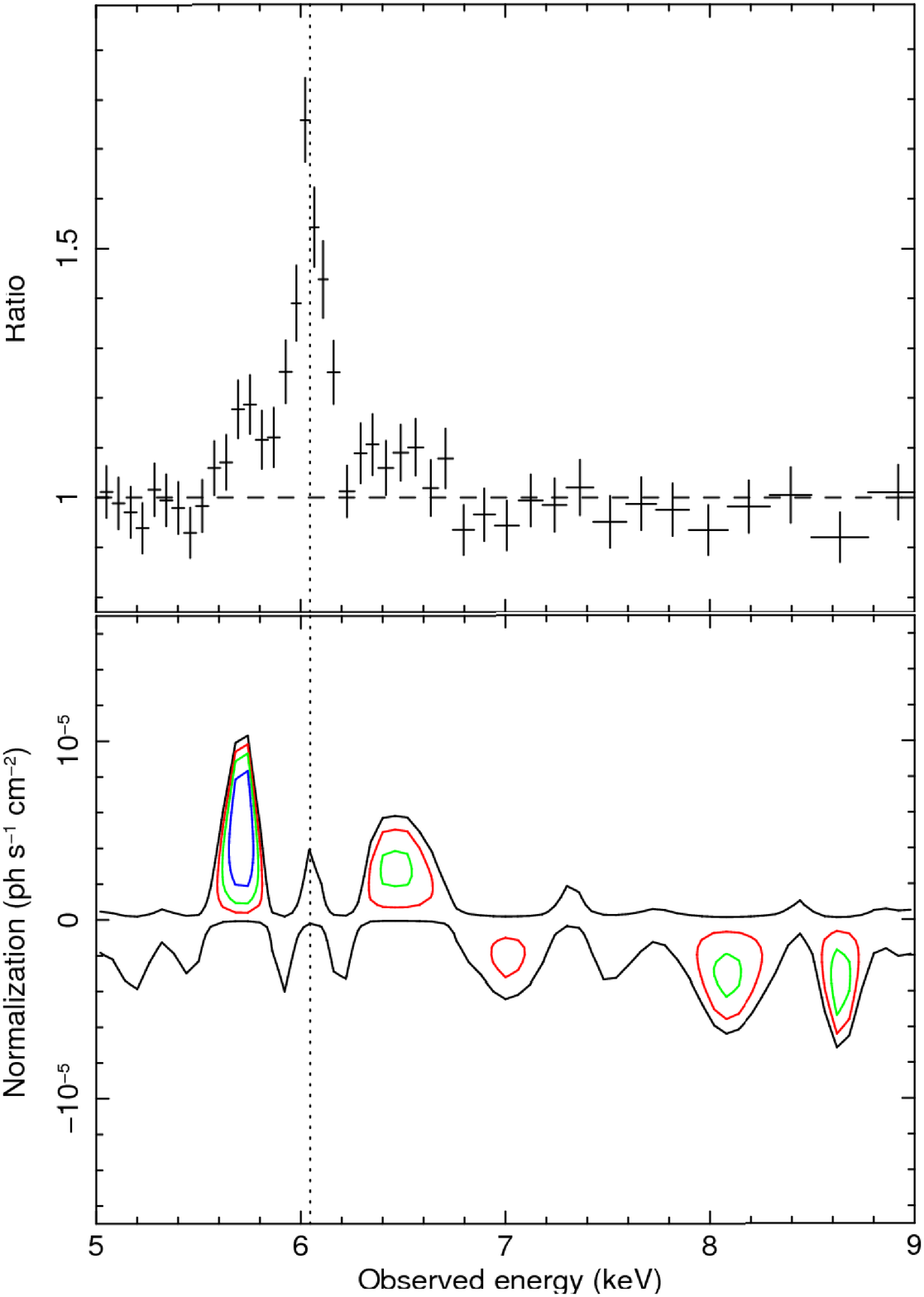}
   \caption{Suzaku XIS-FI spectra of 3C~382 (\emph{left}) and 3C~445 (\emph{right}) zoomed in the 5--9~keV band to emphasize the Fe K complex. \emph{Upper panel:} ratio against a power-law continuum (with a neutral absorption component for 3C~445). For plotting purposes only, data have been further grouped with XSPEC to reach a S/N in each energy bin of 40 and 20 for 3C~382 and 3C~445, respectively. \emph{Lower panel:} energy-intensity contours with respect to the baseline model described in Table~2 (see \S~3.1 and \S~3.2 for more details).}
     \end{figure*}

\clearpage

\begin{deluxetable}{lccccccc}
\tabletypesize{\footnotesize}
\tablecaption{List of sources and Suzaku XIS-FI observations.}
\tablewidth{0pt}
\tablehead{\colhead{Source} & \colhead{$z$} & \colhead{$N_{H,Gal}$} & \colhead{OBSID} & \colhead{Date} & \colhead{Net Expo} & \colhead{Flux} & \colhead{Source/Bkgd}\\ 
& & \colhead{\scriptsize{($10^{20}$~\textbf{cm}$^{-2}$)}} & & & \colhead{\scriptsize{(ks)}} & \colhead{\scriptsize{($10^{-11}$~erg~s$^{-1}$~cm$^{-2}$)}} & \colhead{\scriptsize{($10^3$~cts)}}}
\startdata
3C~111 & 0.0485 & 30.0 & 703034010\tablenotemark{a} & 2008-08-22 & 109 & 1.3 & 9.2/0.6\\
3C~390.3 & 0.0561 & 3.8 & 702125010\tablenotemark{a} & 2007-04-27 & 85 & 2.0 & 13.3/0.5\\
3C~120a & 0.0330 & 11.0 & 700001010\tablenotemark{b} & 2006-02-09 & 42 & 2.9 & 13.8/0.3\\
3C~120b\tablenotemark{c} & 0.0330 & 11.0 & 700001020\tablenotemark{b} & 2006-02-16 & 42 & 2.6 & 12.5/0.3\\
3C~120b\tablenotemark{c} & 0.0330 & 11.0 & 700001030\tablenotemark{b} & 2006-02-23 & 41 & 2.6 & 12.2/0.3\\
3C~120b\tablenotemark{c} & 0.0330 & 11.0 & 702125010\tablenotemark{b} & 2006-03-02 & 41 & 2.5 & 11.7/0.3\\ 
3C~382 & 0.0579 & 7.4 & 701060010\tablenotemark{a} & 2006-12-14 & 116 & 2.5 & 21.2/0.5\\
3C~445 & 0.0562 & 4.8 & 702056010\tablenotemark{a} & 2007-05-25 & 108 & 0.7 & 13.8/0.3\\
\enddata
\tablecomments{Column 1: source name. Column 2: cosmological red-shift. Column 3: neutral Galactic absorption column density. Column 4: observation ID. Column 5: starting date of the observation (in year-month-day). Column 6: net exposure for each XIS. Column 7: flux in the 4--10~keV band. Column 8: total source/background counts in the 7--10~keV band.}
\tablenotetext{a}{For the XIS~0 and XIS~3 cameras combined.}
\tablenotetext{b}{For the XIS~0, XIS~2 and XIS~3 cameras combined.}
\tablenotetext{c}{These observations have been added together in 3C~120b.}
\end{deluxetable}

\clearpage

\begin{deluxetable}{llccllc}
\tabletypesize{\footnotesize}
\tablecaption{Best fit baseline models in the 3.5--10.5~keV band.}
\tablewidth{0pt}
\tablehead{ Source & \colhead{$\Gamma$} & \colhead{$N_{H}$} & \colhead{E} & \colhead{$\sigma$} & \colhead{EW} & \colhead{$\chi^2/\nu$} \\ 
 & & \colhead{($10^{22}$~cm$^{-2}$)} & \colhead{(keV)} & \colhead{(eV)} & \colhead{(eV)} & }
\startdata
3C~111 & $1.47^{+0.02}_{-0.04}$ & $\dots$ & $6.40\pm0.01$ & $110^{+25}_{-19}$ & $86\pm16$ & 412/427\\[2pt]
3C~390.3 & $1.58\pm0.01$ & $\dots$ & $6.42\pm0.01$ & $120^{+25}_{-20}$ & $68\pm14$ & 466/450\\[2pt] 
3C~120a & $1.75\pm0.01$ & $\dots$ & $6.40\pm0.02$ & $90^{+25}_{-31}$ & $68\pm13$ & 1386/1393\\[2pt]
3C~120b & $1.67\pm0.01$ & $\dots$ & $6.38\pm0.01$ & $130^{+13}_{-16}$ & $90\pm10$ & 1743/1707\\[2pt]
 & & & $6.94\pm0.03$ & $83^{+32}_{-28}$ & $24\pm8$ &\\[2pt]
3C~382 & $1.75\pm0.01$ & $\dots$ & $6.40\pm0.02$ & $120\pm20$ & $60\pm11$ & 1490/1516\\[2pt]
 & & & $6.91\pm0.02$ & 10\tablenotemark{a} & $16\pm8$ &\\[2pt]
3C~445 & $1.64\pm0.04$ & $19\pm4$ & $6.38\pm0.01$ & $50\pm20$ & $133^{+22}_{-20}$ & 416/391\\
\enddata
\tablecomments{Column 1: source name. Column 2: power-law photon index. Column 3: equivalent Hydrogen column density due to neutral absorption intrinsic to the source, if present. Column 4: rest-frame energy of the Gaussian emission line. Column 5: line width. Column 6: equivalent width. Column 7: ratio between best fit $\chi^2$ and degrees of freedom. Errors are at the 1$\sigma$ level.}
\tablenotetext{a}{Parameter held fix during the fit.}
\end{deluxetable}

\clearpage

\begin{deluxetable}{lcccccccc}
\tabletypesize{\footnotesize}
\tablecaption{Absorption line parameters.}
\tablewidth{0pt}
\tablehead{ Source & \colhead{ID} & \colhead{E} & \colhead{$\sigma$} & \colhead{EW} & \colhead{$\Delta\chi^2 / \Delta\nu$} & \colhead{$\chi^2/\nu$} & \colhead{F-test} & \colhead{MC} \\ 
 & & \colhead{(keV)}& \colhead{(eV)} & \colhead{(eV)} & & & & }
\startdata
3C~111 & Ly$\alpha$ & $7.26(6.92)^{+0.03}_{-0.03}$ & 10\tablenotemark{a} & $-31\pm15$ & 13/2 & 359/422 & 99.9\% & 99\%\\[0.5pt]
 & Ly$\beta$-Ly$\gamma$-Ly$\delta$ & $8.69(8.29)^{+0.13}_{-0.08}$ & $390^{+270}_{-70}$ & $-154\pm80$ & 40/3 & & $\ge$99.9\% & $\ge$99.9\%\\[7.5pt]
3C~390.3 & Ly$\alpha$ &  $8.11(7.68)^{+0.04}_{-0.04}$ & 10\tablenotemark{a} & $-32\pm16$ & 14.6/2 & 451/448 & 99.9\% & 99.5\%\\[7.5pt]
3C~120a & $\dots$ & $\equiv$7.25\tablenotemark{a} & 10\tablenotemark{a} & $>-29$\tablenotemark{b} & $\dots$ & $\dots$ & $\dots$ & $\dots$\\[0.5pt]
 & $\dots$ & $\equiv$7.54\tablenotemark{a} & 10\tablenotemark{a} & $>-32$\tablenotemark{b} & $\dots$ & $\dots$ & $\dots$ & $\dots$\\[0.5pt]
 & $\dots$ & $\equiv$8.76\tablenotemark{a} & 360\tablenotemark{a} & $>-160$\tablenotemark{b} & $\dots$ & $\dots$ & $\dots$ & $\dots$\\[7.5pt]
3C~120b & He$\alpha$ & $7.25(7.02)^{+0.03}_{-0.03}$ & 10\tablenotemark{a} & $-10\pm5$ & 9.4/2 & 1705/1700 & 99\% & 91\%\\[0.5pt]
 & Ly$\alpha$ & $7.54(7.30)^{+0.04}_{-0.04}$ & 10\tablenotemark{a} & $-12\pm6$ & 10/2 & & 99.3\% & 92\%\\[0.5pt]
 & He$\beta$-Ly$\beta$ & $8.76(8.48)^{+0.12}_{-0.12}$ & $360^{+160}_{-120}$ & $-50\pm13$ & 18/3 & & 99.9\% & 99.8\%\\[7.5pt]
3C~382 & $\dots$ & $\equiv$8\tablenotemark{a} & 10\tablenotemark{a} & $>-20$\tablenotemark{b} & $\dots$ & $\dots$ & $\dots$ & $\dots$\\[7.5pt]
3C~445 & $\dots$ & $\equiv$8\tablenotemark{a} & 10\tablenotemark{a} & $>-45$\tablenotemark{b} & $\dots$ & $\dots$ & $\dots$ & $\dots$\\
\enddata
\tablecomments{Column 1: source name. Column 2: absorption line identification, He$\alpha$/He$\beta$ refer to K-shell transitions from Fe XXV, Ly$\alpha$/Ly$\beta$/Ly$\gamma$/Ly$\delta$ refer to the Fe XXVI Lyman series and the ``-'' indicates a possible line blending (see text for more details). Column 3: absorption line rest-frame (observer frame) energy. Column 4: line width. Column 5: line equivalent width. Column 6: $\chi^2$ improvement adding the absorption line to the baseline model reported in Table~2 and relative number of new parameters. Column 7: ratio between the best fit $\chi^2$ and degrees of freedom after the inclusion of the Gaussian absorption lines. Column 8: detection confidence level from the F-test. Column 9: detection confidence level from extensive Monte Carlo simulations. Errors are at the 1$\sigma$ level.}
\tablenotetext{a}{Parameter held fix during the fit.}
\tablenotetext{b}{Equivalent width lower limit at the 90\% level.}
\end{deluxetable}

\clearpage

\begin{deluxetable}{lcccc}
\tabletypesize{\footnotesize}
\tablecaption{Best fit Xstar photo-ionization models for the observations with detected UFOs.}
\tablewidth{0pt}
\tablehead{ Source & \colhead{log$\xi$} & \colhead{$N_H$} & \colhead{$v_{out}$} & \colhead{$\chi^2/\nu$}\\ 
 & \colhead{(erg~s$^{-1}$~cm)}& \colhead{($10^{22}$~cm$^{-2}$)} & \colhead{(c)} & }
\startdata
3C~111 & $5.0\pm0.3$ & $>20$\tablenotemark{a} & $+0.041\pm0.003$ & 390/424\\[2.5pt]
3C~390.3 & $5.6^{+0.2}_{-0.8}$ & $>3$\tablenotemark{a} & $+0.146\pm0.004$ & 452/447\\[2.5pt] 
3C~120b & $3.8\pm0.2$ & $1.1^{+0.5}_{-0.4}$ & $+0.076\pm0.003$ & 1731/1704\\[2.5pt]
\enddata
\tablecomments{Column 1: source name. Column 2: ionization parameter. Column 3: equivalent Hydrogen column density of the ionized absorber. Column 4: blue-shifted (outflow) velocity. Column 5: ratio between the best fit $\chi^2$ and degrees of freedom after the inclusion of the Xstar model. Errors are at the 1$\sigma$ level.}
\tablenotetext{a}{{Lower limit at the 90\% level.}}
\end{deluxetable}

\clearpage

\begin{deluxetable}{lcccc}
\tabletypesize{\footnotesize}
\tablecaption{Consistency checks for the blue-shifted Fe K absorption lines.}
\tablewidth{0pt}
\tablehead{ Source & \colhead{Inst} & \colhead{E} & \colhead{$\sigma$} & \colhead{EW}\\ 
 & & \colhead{(keV)}& \colhead{(eV)} & \colhead{(eV)}}
\startdata
3C~111 & XIS~0 & $7.27\pm0.04$ & 10 & $-36\pm22$\\
       &       & $8.80\pm0.14$ & $380^{+540}_{-130}$ & $-140\pm40$\\[2.5pt]
       & XIS~3 & $7.24\pm0.02$ & 10 & $-34\pm19$\\
       &       & $8.61\pm0.09$ & $390^{+180}_{-110}$ & $-170^{+30}_{-40}$\\[2.5pt]
       & XIS-BI & $\equiv$7.26\tablenotemark{a} & 10 & $>-60$\tablenotemark{b}\\
       &        & $\equiv$8.69\tablenotemark{a} & $\equiv$390\tablenotemark{a} & $>-190$\tablenotemark{b}\\[2.5pt]
\hline\\[-5pt]
3C~390.3 & XIS~0 & $8.06\pm0.07$ & 10 & $>-50$\tablenotemark{b}\\[2.5pt]
         & XIS~3 & $8.11\pm0.03$ & 10 & $-32\pm19$\\[2.5pt]
         & XIS-BI & $\equiv$8.11\tablenotemark{a} & 10 & $>-50$\tablenotemark{b}\\[2.5pt]
\hline\\[-5pt]
3C~120b & XIS~0 & $7.24^{+0.10}_{-0.16}$ & 10 & $>-40$\tablenotemark{b}\\
        &       & $7.58\pm0.08$ & 10 & $>-50$\tablenotemark{b}\\
        &       & $8.83\pm0.18$ & $285^{+220}_{-130}$ & $-70\pm40$\\[2.5pt]
        & XIS~2 & $7.25\pm0.05$ & 10 & $>-46$\tablenotemark{b}\\
        &       & $7.51\pm0.07$ & 10 & $>-42$\tablenotemark{b}\\
        &       & $8.79^{+0.25}_{-0.29}$ & $312^{+339}_{-264}$ & $-55\pm33$\\[2.5pt]
        & XIS~3 & $7.24\pm0.05$ & 10 & $>-43$\tablenotemark{b}\\
        &       & $7.65\pm0.10$ & 10 & $>-48$\tablenotemark{b}\\
        &       & $\equiv$8.76\tablenotemark{a} & $\equiv$360\tablenotemark{a} & $>-70$\tablenotemark{b}\\[2.5pt]
        & XIS-BI & $\equiv$7.25\tablenotemark{a} & 10 & $>-40$\tablenotemark{b}\\
        &        & $\equiv$7.54\tablenotemark{a} & 10 & $>-48$\tablenotemark{b}\\
        &        & $\equiv$8.76\tablenotemark{a} & $\equiv$360\tablenotemark{a} & $>-71$\tablenotemark{b}\\
\enddata
\tablecomments{\scriptsize{Column 1: source name. Column 2: Suzaku instrument; broad stands for broad band fit, using both XIS-FI and PI. Column 3: Absorption line rest-frame energy. Column 4: line width. Column 5: line equivalent width. Errors are at the 1$\sigma$ level.}}
\tablenotetext{a}{\scriptsize{Parameter held fix during the fit.}}
\tablenotetext{b}{\scriptsize{Equivalent width lower limit at the 90\% level.}}
\end{deluxetable}

\end{document}